\documentclass[12pt]{iopart}

\usepackage{ulem}
\usepackage{newtxtext}
\usepackage{color}
\usepackage{comment}
\usepackage{cite}
\usepackage{amssymb}
\usepackage{amsthm}
\usepackage{iopams}

\theoremstyle{definition}
\newtheorem{Def}{Definition}[section]

\newtheorem{thm}[Def]{Theorem}
\newtheorem{rem}[Def]{Remark}

\usepackage{graphicx}

\begin{document}

\title
[Lax pairs for delay soliton equations and their reductions to delay Painlev\'e equations]
{Construction of the Lax pairs for the delay Lotka-Volterra and delay Toda lattice equations and their reductions to delay Painlev\'e equations}
\author{Hiroshi Matsuoka$^1$, Kenta Nakata$^1$ and Ken-ichi Maruno$^2$$^\dagger$}
\address{$^1$~Department of Pure and Applied Mathematics, School of Fundamental Science and Engineering, Waseda University, 3-4-1 Okubo, Shinjuku-ku, Tokyo 169-8555, Japan}
\address{$^2$~Department of Applied Mathematics, Faculty of Science and Engineering, Waseda University, 3-4-1 Okubo, Shinjuku-ku, Tokyo 169-8555, Japan}
\address{$^2$~Corresponding author}
\ead{matsuoka8@akane.waseda.jp kennakaxx@akane.waseda.jp kmaruno@waseda.jp}
\vspace{10pt}
\begin{indented}
\item[]{\today}
\end{indented}
\begin{abstract}
    The delay Lotka-Volterra and delay Toda lattice equations are delay-differential extensions of the well-known soliton equations, the Lotka-Volterra and Toda lattice equations, respectively.
    This paper investigates integrability properties of the delay Lotka-Volterra and delay Toda lattice equations,
    and study the relationships to the already known delay Painlev\'e equations.
    First, B\"{a}cklund transformations, Lax pairs and an infinite number of conserved quantities of these delay soliton equations are constructed.
    Then, applying spatial $2$-periodic reductions to them, we show the known delay Painlev\'e equations are derived.
    Using these reductions, we construct the $N$-soliton-type determinant solutions of the autonomous versions of delay Painlev\'e equations, and the Casorati determinant solution of a higher order analogue of the discrete Painlev\'e I\hspace{-1pt}I equation.
\end{abstract}
\noindent{\it Keywords\/}:\ \ discrete-time delay soliton equations, delay soliton equations, delay Painlev\'e equations, B\"{a}cklund transformations, Lax pairs, conserved quantities, determinant solutions

\submitto{\NL}
\footnotetext{Corresponding author}

\begin{section}{Introduction}
\label{sec_intro}

A delay-differential equation is defined as an equation which includes both derivatives and shifts with respect to the same independent variable.
From the viewpoint of applied mathematics, various traffic flow and infectious disease phenomena, for example, are modelled by delay-differential equations~\cite{delaybook,Culshaw,Xiao,Tutiya}.

Delay-differential equations have also been studied from mathematical viewpoints.
In the research field of integrable systems,
integrable delay-differential analogues of Painlev\'e equations were proposed and have been studied from various approaches, such as singularity confinement, bilinear forms, algebraic entropy, and exact solutions~\cite{Quispel,Levi,Gram,Ramani,Joshi1,Joshi2,Carstea,Viallet,Viallet2,Halburd,Berntson,Stokes}.
Recently, Gibbons \textit{et al} pointed out that Masur-Veech volumes of moduli spaces of quadratic differentials can be computed by using the delay Painlev\'e I equation~\cite{Gibbons}.
Thus, delay Painlev\'e equations are expected to play important roles not only in integrable systems but also in other areas of mathematics.

On the other hand, several integrable delay-differential soliton equations were proposed and considered~\cite{Ablowitz,Tutiya2,Sekiguchi,Tsunematsu}.
Recently, a systematic method for constructing delay analogues of \sout{the} already known soliton equations and their discrete analogues was proposed~\cite{Nakata1,Nakata2}.
Such delay soliton equations have $N$-soliton solutions and are reduced to \sout{the} already known soliton equations in the small limit of delay parameters.
While studies of delay soliton equations were applied to the box and ball system~\cite{Nakata3}, the properties as integrable systems have not yet been thoroughly investigated.
In this paper, we investigate the integrability of delay soliton equations and clarify the relationships between delay Painlev\'{e} equations and delay soliton equations.

We first show a method for constructing Lax pairs of delay soliton equations and their discrete analogues. This method is discussed and demonstrated by using the delay Lotka-Volterra (LV) and the delay Toda lattice equations~\cite{Nakata1} as examples.
Using their Lax pairs, conserved quantities of them are also derived.
These conserved quantities are those of a system of ordinary difference equations, namely the delay soliton equation with a periodic boundary condition in the spatial variable, and depend only on the time variable.
Note that \textit{a discrete analogue of a delay soliton equation} is a higher order partial difference equation that involves delay parameters as shifts of the independent variables and leads to an already known delay soliton equation by the continuum limit.
%They are written by using the time-ordered product, which is common in quantum field theory.

Then, we discuss relationships between delay soliton equations and delay Painlev\'e equations.
It is well known that Painlev\'e equations can be derived by similarity reductions of soliton equations~\cite{Ablowitz2}.
In addition, 
Nijhoff \textit{et al}~\cite{Nijhoff1,Nijhoff2,Nijhoff3}, Kajiwara \textit{et al}~\cite{Kajiwara2} and Grammaticos \textit{et al}~\cite{Gram2,Gram3} 
derived discrete Painlev\'e equations by reductions of discrete soliton equations.
According to these studies, we can naturally expect that reductions of delay soliton equations lead to delay Painlev\'e equations.
In this paper, we apply spatial $2$-periodic reductions to the delay LV and delay Toda lattice equations, and then derive the delay Painlev\'e I\hspace{-1pt}I and I\hspace{-1pt}I\hspace{-1pt}I equations~\cite{Gram}.

By imposing these reductions on determinant solutions of the delay LV and delay Toda lattice equations, we construct the $N$-soliton-type determinant solutions of the autonomous cases of the delay Painlev\'e I\hspace{-1pt}I and I\hspace{-1pt}I\hspace{-1pt}I equations.
These are solutions of autonomous ordinary delay-differential equations, with arbitrary size determinant structures.
These examples exist because of the delay parameters included in the equations.
Berntson~\cite{Berntson} constructed special solutions of autonomous cases of delay Painlev\'e equations, in parallel with well-known elliptic solutions of autonomous limits of Painlev\'e equations.
In particular, the 3-soliton-type solutions of the autonomous delay Painlev\'e I\hspace{-1pt}I and I\hspace{-1pt}I\hspace{-1pt}I equations were obtained by Hirota's direct method.
Using the $N$-soliton-type determinant solutions obtained in this paper, we can easily reconstruct this result.
Finally, we construct the Casorati determinant solution of the discrete analogue of the delay Painlev\'e I\hspace{-1pt}I equation by using Pl\"ucker relations. Note that \textit{a discrete analogue of a delay Painlev\'{e} equation} is a higher order ordinary difference equation whose order is determined by an integer-valued delay parameter and that leads to an already known delay Painlev\'{e} equation in the continuum limit.

This paper is structured as follows.
In section \ref{sec_disLV} and \ref{sec_contiLV}, we first construct B\"{a}cklund transformations, Lax pairs, and conserved quantities of the delay LV equation and its discrete analogue.
In section \ref{sec_disTL} and \ref{sec_contiTL}, those of the delay Toda lattice equation and its discrete analogue are obtained.
In section \ref{sec_reduction}, we consider spatial $2$-periodic reductions of the delay LV and delay Toda lattice equations, leading to the delay Painlev\'e I\hspace{-1pt}I and I\hspace{-1pt}I\hspace{-1pt}I equations.
The discrete analogues of the delay Painlev\'e I\hspace{-1pt}I and I\hspace{-1pt}I\hspace{-1pt}I equations are obtained in parallel.
In section \ref{sec_detsol}, we derive several determinant solutions of the autonomous delay Painlev\'e I\hspace{-1pt}I and I\hspace{-1pt}I\hspace{-1pt}I equations and the discrete analogue of the delay Painlev\'e I\hspace{-1pt}I equation using the above reductions or Pl\"ucker relations.
In section \ref{sec_con}, we discuss the results and give conclusions.

\end{section}

\begin{section}{The Lax pair and conserved quantities of the discrete-time delay Lotka-Volterra equation}
\label{sec_disLV}

In this section, we construct the B\"{a}cklund transformation, Lax pair, and conserved quantities of the discrete-time delay LV equation, which can be considered as a higher order discrete LV equation, before considering the delay LV equation.

\begin{subsection}{Construction of the discrete-time delay Lotka-Volterra equation}

We start from the discrete analogue of a generalized Toda equation (DAGTE), which was introduced by Hirota~\cite{Hirota1,Miwa1}:
\begin{eqnarray}
    \{Z_{1}\exp\left({D_{1}}\right)+Z_{2}\exp\left({D_{2}}\right)+Z_{3}\exp\left({D_{3}}\right)\}f\cdot f=0\label{DAGTE},
\end{eqnarray}
where $f=f(n,m,k)$, $Z_{1}+Z_{2}+Z_{3}=0$ and $D_{1}$, $D_{2}$ and $D_{3} $ are linear sums of Hirota's D operators.
Here Hirota's D-operators are defined by
\begin{eqnarray}
    \fl D_{t}^{l}a(t)\cdot b(t)=\left.\left(\frac{\partial}{\partial t}-\frac{\partial}{\partial s}\right)^{l}a(t)b(s)\right|_{s=t},\quad \exp\left(D_{m}\right)a_{m}\cdot b_{m}=a_{m+1}b_{m-1}.
\end{eqnarray}
The B\"{a}cklund transformation of equation (\ref{DAGTE}) was obtained as follows~\cite{Hirota1}:
\begin{eqnarray}
    \fl \left(\lambda_{1}\exp\left(\frac{D_{1}+D_{3}}{2}\right)-\exp\left(-\frac{D_{1}+D_{3}}{2}\right)-\mu_{1}\exp\left(\frac{D_{1}+2D_{2}-D_{3}}{2}\right)\right)f\cdot g=0,\label{DAGTEBT1}\\
    \fl \left(\lambda_{2}\exp\left(\frac{D_{2}+D_{3}}{2}\right)+\exp\left(-\frac{D_{2}+D_{3}}{2}\right)+\mu_{2}\exp\left(\frac{D_{2}+2D_{1}-D_{3}}{2}\right)\right)f\cdot g=0,\label{DAGTEBT2}
\end{eqnarray}
where $g$ is a new solution of equation (\ref{DAGTE}) obtained from the solution $f$, and $\lambda_{1},\ \lambda_{2}$ are arbitrary constants, and $\mu_{1},\ \mu_{2}$ satisfy $\mu_{1}Z_{1}+\mu_{2}Z_{2}=0$.

Now, we consider a reduction of the DAGTE (\ref{DAGTE}).
By setting
\begin{eqnarray}
    \eqalign{
    Z_{1}=\delta,\quad Z_{2}=1,\quad Z_{3}=-(1+\delta),\\
    D_{1}=\frac{(1-\alpha)D_{m}-(3+\beta)D_{n}}{2},\\
    D_{2}=\frac{-(1-\alpha)D_{m}+(1+\beta)D_{n}}{2},\\
    D_{3}=\frac{(1+\alpha)D_{m}+(1+\beta)D_{n}}{2},}\label{reductionLV}
\end{eqnarray}
equation (\ref{DAGTE}) is reduced to the bilinear form of the discrete-time delay LV equation~\cite{Nakata1,Nakata3}:
\begin{eqnarray}
    (1+\delta)f_{n+\beta}^{m+1+\alpha}f_{n-1}^{m}-\delta f_{n+1+\beta}^{m+\alpha}f_{n-2}^{m+1}-f_{n+\beta}^{m+\alpha}f_{n-1}^{m+1}=0,\label{dlydisLVb}
\end{eqnarray}
where $f_{n}^{m}=f(n,m,0)$ and $\alpha, \beta$ are constant integers.
The nonlinear form of the discrete-time delay LV equation 
\begin{eqnarray}
    \frac{u_{n+\beta}^{m+1+\alpha}u_{n-1}^{m}}{u_{n+\beta}^{m+\alpha}u_{n-1}^{m+1}}=\frac{(1+\delta u_{n+1+\beta}^{m+\alpha})(1+\delta u_{n-2}^{m+1})}{(1+\delta u_{n+\beta}^{m+\alpha})(1+\delta u_{n-1}^{m+1})}\label{dlydisLVn}
\end{eqnarray}
is obtained through the dependent variable transformation
\begin{eqnarray}
    u_{n}^{m}=\frac{f_{n+1+\beta}^{m+\alpha}f_{n-2}^{m+1}}{f_{n+\beta}^{m+\alpha}f_{n-1}^{m+1}}.\label{dlydisLVtrans}
\end{eqnarray}
Equation (\ref{dlydisLVn}) leads to the division of the following discrete LV equations when $\alpha=\beta=0$:
\begin{eqnarray}
    \frac{u_{n}^{m+1}}{u_{n}^{m}}
    =\frac{1+\delta u_{n+1}^{m}}{1+\delta u_{n-1}^{m+1}},\qquad
    \frac{u_{n-1}^{m+1}}{u_{n-1}^{m}}
    =\frac{1+\delta u_{n}^{m}}{1+\delta u_{n-2}^{m+1}}.\label{disLVn}
\end{eqnarray}
By introducing the alternative dependent variables
\begin{eqnarray}
    R_{n}^{m}
    =\frac{f_{n-1}^{m}f_{n}^{m+1}}{f_{n}^{m}f_{n-1}^{m+1}},\qquad
    %=\frac{1}{(1+\delta)}\prod_{k=1}^{\infty}\frac{1+\delta u_{n+1-k(1+\beta)}^{m-k\alpha}}{1+\delta u_{n-k(1+\beta)}^{m-k\alpha}},\\
    S_{n}^{m}
    =\delta\frac{f_{n-1}^{m}f_{n+2+\beta}^{m+\alpha}}{f_{n}^{m}f_{n+1+\beta}^{m+\alpha}},
    %=\frac{\delta u_{n+1}^{m}}{(1+\delta)}\prod_{k=1}^{\infty}\frac{1+\delta u_{n+1-k(1+\beta)}^{m-k\alpha}}{1+\delta u_{n-k(1+\beta)}^{m-k\alpha}},
    \label{dlydisLVRS}
\end{eqnarray}
equation (\ref{dlydisLVb}) is transformed into
\begin{eqnarray}
    \eqalign{
    R_{n}^{m}S_{n}^{m+1}=R_{n+2+\beta}^{m+\alpha}S_{n}^{m},\\
    R_{n+1+\beta}^{m+\alpha}-R_{n}^{m}=S_{n}^{m}-S_{n-1}^{m+1}.}\label{dlydisLVn_2var}
\end{eqnarray}

\end{subsection}

\begin{subsection}{A B\"{a}cklund transformation and Lax pair of the discrete-time delay Lotka-Volterra equation}

By applying the reduction condition (\ref{reductionLV}) to equations (\ref{DAGTEBT1}) and (\ref{DAGTEBT2}) with the setting $\mu_{1} = \mu$, $\mu_{2} = -\mu \delta$, where $\mu$ is a constant, a B\"{a}cklund transformation of the discrete-time delay LV equation (\ref{dlydisLVb}) is obtained as follows:

\begin{eqnarray}
    \lambda_{1}f_{n}^{m+1}g_{n+1}^{m}-f_{n+1}^{m}g_{n}^{m+1}-\mu f_{n}^{m}g_{n+1}^{m+1}=0,\label{dlydisLVBT1}\\
    \lambda_{2}f_{n+1+\beta}^{m+\alpha}g_{n}^{m}+f_{n}^{m}g_{n+1+\beta}^{m+\alpha}-\mu\delta f_{n-1}^{m}g_{n+2+\beta}^{m+\alpha}=0.\label{dlydisLVBT2}
\end{eqnarray}
\begin{rem}
Equation (\ref{DAGTE}) is invariant under the transformations
\begin{eqnarray*}
    (Z_{1}, D_{1}, Z_{2}, D_{2}, Z_{3}, D_{3})\to
    (Z_{\sigma(1)}, \pm D_{\sigma(1)}, Z_{\sigma(2)}, \pm D_{\sigma(2)}, Z_{\sigma(3)}, \pm D_{\sigma(3)}),
\end{eqnarray*}
where $\sigma$ is a permutation of $(1,2,3)$.
Therefore, equations (\ref{DAGTEBT1}) and (\ref{DAGTEBT2}) with these transformations are also B\"{a}cklund transformations of the DAGTE.
For example, applying the transformation
\begin{eqnarray*}
    (Z_{1}, D_{1}, Z_{2}, D_{2}, Z_{3}, D_{3})\to
    (Z_{1}, D_{1}, Z_{3}, -D_{3}, Z_{2}, -D_{2})
\end{eqnarray*}
to the reduction condition (\ref{reductionLV}), we obtain a new B\"{a}cklund transformation of the discrete-time delay LV equation (\ref{dlydisLVb}):
\begin{eqnarray}
    \lambda_{1}f_{n-1}^{m+1}g_{n+1+\beta}^{m+\alpha}-f_{n+1+\beta}^{m+\alpha}g_{n-1}^{m+1}+\mu(1+\delta) f_{n-1}^{m}g_{n+1+\beta}^{m+1+\alpha}=0,\label{dlydisLVBT1_alt}\\
    \lambda_{2}f_{n}^{m}g_{n+1+\beta}^{m+\alpha}+f_{n+1+\beta}^{m+\alpha}g_{n}^{m}-\mu\delta f_{n-1}^{m}g_{n+2+\beta}^{m+\alpha}=0.\label{dlydisLVBT2_alt}
\end{eqnarray}
However, this B\"{a}cklund transformation is not appropriate for the construction of the Lax pair as discussed in Remark \ref{rem_dlydisLVLax_alt}.
\end{rem}

Next, we construct a Lax pair of the discrete-time delay LV equation.
Setting $g_{n}^{m}=f_{n}^{m}\psi_{n}^{m}$, we obtain the following equations by (\ref{dlydisLVBT1}) and (\ref{dlydisLVBT2}):
\begin{eqnarray}
    \mu R_{n}^{m}\psi_{n}^{m+1}+\psi_{n-1}^{m+1}=\lambda_{1}\psi_{n}^{m},\label{dlydisLVLax1}\\
    \mu S_{n}^{m}\psi_{n+2+\beta}^{m+\alpha}-\psi_{n+1+\beta}^{m+\alpha}=\lambda_{2}\psi_{n}^{m},\label{dlydisLVLax2}
\end{eqnarray}
where $R_{n}^{m}$ and $S_{n}^{m}$ are defined by (\ref{dlydisLVRS}).
We introduce linear operators $\bar{L_{1}}$ and $\bar{L_{2}}$ by
\begin{eqnarray}
    \bar{L_{1}}(n,m)=\mu R_{n}^{m}+\exp\left(-\frac{\partial}{\partial n}\right),\\
    \bar{L_{2}}(n,m)=\mu S_{n}^{m}\exp\left((2+\beta)\frac{\partial}{\partial n}\right)-\exp\left((1+\beta)\frac{\partial}{\partial n}\right),
\end{eqnarray}
where the shift operator is defined by
\begin{eqnarray}
    \exp\left(a\frac{\partial}{\partial n}\right)f(n)=f(n+a),
\end{eqnarray}
for a constant $a$.
Then equations (\ref{dlydisLVLax1}) and (\ref{dlydisLVLax2}) are rewritten as follows:
\begin{eqnarray}
    \bar{L_{1}}(n,m)\psi_{n}^{m+1}=\lambda_{1}\psi_{n}^{m},\label{dlydisLVLaxL1}\\
    \bar{L_{2}}(n,m)\psi_{n}^{m+\alpha}=\lambda_{2}\psi_{n}^{m}.\label{dlydisLVLaxL2}
\end{eqnarray}
The compatibility condition of equations (\ref{dlydisLVLaxL1}) and (\ref{dlydisLVLaxL2}) is
\begin{eqnarray}
    \bar{L_{1}}(n,m)\bar{L_{2}}(n,m+1)
    =\bar{L_{2}}(n,m)\bar{L_{1}}(n,m+\alpha),\label{dlydisLVcompa}
\end{eqnarray}
and this equation leads to the discrete-time delay LV equation (\ref{dlydisLVn}) or (\ref{dlydisLVn_2var}).
\begin{rem}
When $\alpha=\beta=0$, equations (\ref{dlydisLVLax1}) and (\ref{dlydisLVLax2}) lead to the following equation~\cite{Hirota2}:
\begin{eqnarray}
    \frac{\mu(1+\delta u_{n}^{m})}{(1+\delta)}\psi_{n}^{m+1}+\psi_{n-1}^{m+1}=\lambda_{1}\psi_{n}^{m},\label{disLVLax1}\\
    \mu\delta u_{n+1}^{m}\frac{(1+\delta u_{n}^{m})}{(1+\delta)}\psi_{n+2}^{m}-\psi_{n+1}^{m}=\lambda_{2}\psi_{n}^{m}.\label{disLVLax}
\end{eqnarray}
The compatibility condition of these equations leads to the discrete LV equation (\ref{disLVn}).
\end{rem}

\end{subsection}

\begin{subsection}{The Lax pair and conserved quantities of the discrete-time delay Lotka-Volterra equation under the periodic boundary condition}

Now, we introduce the $N$-periodic boundary condition $\psi_{n+N}^{m}=\psi_{n}^{m}$ in the space variable $n$.
By defining 
\begin{eqnarray}
    \hat{\psi}(m)=(\psi_{0}^{m},\psi_{1}^{m},\psi_{2}^{m},\cdots,\psi_{N-1}^{m})^{T},
\end{eqnarray}
equations (\ref{dlydisLVLaxL1}) and (\ref{dlydisLVLaxL2}) in the case of $n=0,1,\cdots,N-1$ can be expressed in the following form:
\begin{eqnarray}
    \bar{\mathcal{L}}_{1}(m)\hat{\psi}(m+1)=\lambda_{1}\hat{\psi}(m),\label{dlydisLVLaxMatrix1}\\
    \bar{\mathcal{L}}_{2}(m)\hat{\psi}(m+\alpha)=\lambda_{2}\hat{\psi}(m)\label{dlydisLVLaxMatrix2},
\end{eqnarray}
where $\bar{\mathcal{L}}_{1}(m)$ and $\bar{\mathcal{L}}_{2}(m)$ are described by matrices.
For example, they are defined as follows in the case of $\beta=2$:
\begin{eqnarray}
\fl\bar{\mathcal{L}}_{1}(m)=
\left(
\begin{array}{ccccc}
\mu R_{0}^{m} & 0             & 0             & \cdots & 1 \\
1             & \mu R_{1}^{m} & 0             & \cdots & 0 \\
0             & 1             & \mu R_{2}^{m} & \cdots & 0 \\
\vdots        & \vdots        & \ddots        & \ddots & \vdots \\
0             & 0             & \cdots        & 1      & \mu R_{N-1}^{m} \\
\end{array}
\right),\label{dlydisLVLaxMatrixL1}\\
\fl\bar{\mathcal{L}}_{2}(m)=
\left(
\begin{array}{cccccccc}
0      & 0      & 0      & -1     & \mu S_{0}^{m} & 0             & \cdots & 0\\
0      & 0      & 0      & 0      & -1            & \mu S_{1}^{m} & \cdots & 0\\
\vdots & \vdots & \vdots & \vdots & \ddots        & \ddots        & \ddots & \vdots\\
0      & 0      & 0      & 0      & \cdots        & 0             & -1     & \mu S_{N-5}^{m}\\
\mu S_{N-4}^{m} & 0 & 0 & 0       & \cdots        & 0             & 0      & -1\\
-1 & \mu S_{N-3}^{m} & 0 & 0      & \cdots        & 0             & 0      & 0\\
0 & -1 & \mu S_{N-2}^{m} & 0      & \cdots        & 0             & 0      & 0\\
0 & 0 & -1 & \mu S_{N-1}^{m}      & \cdots        & 0             & 0      & 0\\
\end{array}
\right).\label{dlydisLVLaxMatrixL2}
\end{eqnarray}
\begin{rem}
\label{rem_dlydisLVLax_alt}
By using the B\"{a}cklund transformation (\ref{dlydisLVBT1_alt}) and (\ref{dlydisLVBT2_alt}) instead of (\ref{dlydisLVBT1}) and (\ref{dlydisLVBT2}), the following Lax pair is obtained:
\begin{eqnarray}
    -\mu(1+\delta u_{n+1}^{m})R_{n}^{m}\psi_{n+1+\beta}^{m+1+\alpha}
    +\psi_{n-1}^{m+1}
    =\lambda_{1}\psi_{n+1+\beta}^{m+\alpha},\label{dlydisLVLax1_alt}\\
    \mu S_{n}^{m}\psi_{n+2+\beta}^{m+\alpha}
    -\psi_{n}^{m}
    =\lambda_{2}\psi_{n+1+\beta}^{m+\alpha}.\label{dlydisLVLax2_alt}
\end{eqnarray}
However, by using the above Lax pair, we cannot construct the matrices $\bar{\mathcal{L}}_{1}(m)$ and $\bar{\mathcal{L}}_{2}(m)$,
because the upper indexes of $\psi$'s in the first and second terms on the left-hand sides in equations  (\ref{dlydisLVLax1_alt}) and (\ref{dlydisLVLax2_alt}) are not equal.
If Lax pairs cannot be written in terms of matrices, it is difficult to compute traces and conserved quantities cannot be easily constructed using the method we use below.
Therefore, it is necessary to select an appropriate B\"{a}cklund transformation to construct the Lax pair.
\end{rem}

Now, by using the above Lax pair, we construct conserved quantities as functions of $m$ under the $N$-periodic boundary condition in the spatial variable $n$.
We consider the case $\alpha\neq0$ and define a new linear operator 
\begin{eqnarray}
\fl\bar{\mathcal{N}}(m)=
\left\{
\begin{array}{ll}
    \bar{\mathcal{L}}_{1}(m)\bar{\mathcal{L}}_{1}(m+1)\cdots \bar{\mathcal{L}}_{1}(m+\alpha-1)\bar{\mathcal{L}}_{2}(m)^{-1} & (\alpha\geq1) \\
    \bar{\mathcal{L}}_{2}(m)\bar{\mathcal{L}}_{1}(m+\alpha)\bar{\mathcal{L}}_{1}(m+\alpha+1)\cdots \bar{\mathcal{L}}_{1}(m-1) & (\alpha\leq-1)
\end{array}
\right..
\label{dlydisLVop0}
\end{eqnarray}
The compatibility condition of equations (\ref{dlydisLVLaxMatrix1}) and (\ref{dlydisLVLaxMatrix2}) is
\begin{eqnarray}
    \bar{\mathcal{L}}_{1}(m)\bar{\mathcal{L}}_{2}(m+1)
    =\bar{\mathcal{L}}_{2}(m)\bar{\mathcal{L}}_{1}(m+\alpha).\label{dlydisLVMatrix_compa1}
\end{eqnarray}
From this condition, the following equation is obtained:
\begin{eqnarray}
    \bar{\mathcal{N}}(m)\bar{\mathcal{L}}_{1}(m)
    =\bar{\mathcal{L}}_{1}(m)\bar{\mathcal{N}}(m+1).\label{dlydisLVMatrix_compa2}
\end{eqnarray}
Defining $\bar{H}_{k}(m)=\mathrm{Tr}\left(\bar{\mathcal{N}}(m)^{k}\right)$, we obtain
\begin{eqnarray}
    \bar{H}_{k}(m+1)
    &=&\mathrm{Tr}\left(\bar{\mathcal{N}}(m+1)^{k}\right)\nonumber\\
    &=&\mathrm{Tr}\left(\bar{\mathcal{L}}_{1}(m)^{-1}\bar{\mathcal{N}}(m)^{k}\bar{\mathcal{L}}_{1}(m)\right)\nonumber\\
    &=&\mathrm{Tr}\left(\bar{\mathcal{N}}(m)^{k}\right)\nonumber\\
    &=&\bar{H}_{k}(m).
\end{eqnarray}
Thus $\bar{H}_{k}(m)\ (k=1,2,\cdots)$ are conserved quantities of the discrete-time delay LV equation.
In the case of $\alpha=0$, we just need to define $\bar{H}_{k}(m):=\mathrm{Tr}\left(\bar{\mathcal{L}_{2}}(m)^{k}\right)$ according to the compatibility condition (\ref{dlydisLVMatrix_compa1}).
On the other hand, when $\alpha\neq0$, equation (\ref{dlydisLVMatrix_compa1}) cannot be used directly to construct conserved quantities.
It is necessary to introduce the operator $\bar{\mathcal{N}}(m)$ to eliminate $\alpha$ in equation (\ref{dlydisLVMatrix_compa1}).
This is one of the problems caused by the delay.

Let us show examples of $\bar{H}_{k}(m)\ (k=1,2,\cdots)$.
Considering the case of $N=3$ and setting $\alpha=2,\ \beta=0,\ \mu=1$, we can calculate $\bar{H}_{1}(m)$ and $\bar{H}_{2}(m)$ as follows:
\begin{eqnarray*}
\fl\bar{H}_{1}(m)=(3+R_{0}^{m}R_{0}^{m+1}S_{2}^{m}+R_{1}^{m}R_{1}^{m+1}S_{0}^{m}+R_{2}^{m}R_{2}^{m+1}S_{1}^{m}+R_{0}^{m+1}S_{0}^{m}S_{2}^{m}+R_{1}^{m+1}S_{1}^{m}S_{0}^{m}\nonumber\\
+R_{2}^{m+1}S_{2}^{m}S_{1}^{m}+R_{0}^{m}S_{1}^{m}S_{2}^{m}+R_{1}^{m}S_{2}^{m}S_{0}^{m}+R_{2}^{m}S_{0}^{m}S_{1}^{m})/(-1+S_{0}^{m}S_{1}^{m}S_{2}^{m}),
\end{eqnarray*}
\begin{eqnarray*}
&\fl \bar{H}_{2}(m)\\
&\fl = (3+2 R_{0}^{m} R_{0}^{m+1} R_{1}^{m+1}+2 R_{0}^{m} R_{1}^{m} R_{1}^{m+1}+2 R_{0}^{m} R_{0}^{m+1} R_{2}^{m}+2 R_{1}^{m} R_{1}^{m+1} R_{2}^{m+1}+2 R_{0}^{m+1} R_{2}^{m} R_{2}^{m+1}\nonumber\\
&\fl+2 R_{1}^{m} R_{2}^{m} R_{2}^{m+1}+2 R_{0}^{m+1} R_{1}^{m+1} S_{0}^{m}+6 R_{1}^{m} R_{1}^{m+1} S_{0}^{m}+2 R_{0}^{m+1} R_{2}^{m} S_{0}^{m}+2 R_{1}^{m} R_{2}^{m} S_{0}^{m}\nonumber\\
&\fl+(R_{1}^{m})^{2} (R_{1}^{m+1})^{2} (S_{0}^{m})^{2}+2 R_{0}^{m} R_{1}^{m+1} S_{1}^{m}+2 R_{0}^{m} R_{2}^{m} S_{1}^{m}+2 R_{1}^{m+1} R_{2}^{m+1} S_{1}^{m}+6 R_{2}^{m} R_{2}^{m+1} S_{1}^{m}\nonumber\\
&\fl+6 R_{1}^{m+1} S_{0}^{m} S_{1}^{m}+6 R_{2}^{m} S_{0}^{m} S_{1}^{m}+2 R_{1}^{m} R_{1}^{m+1} R_{2}^{m} R_{2}^{m+1} S_{0}^{m} S_{1}^{m}+2 R_{1}^{m} (R_{1}^{m+1})^{2} (S_{0}^{m})^{2} S_{1}^{m}\nonumber\\
&\fl+2 R_{1}^{m} R_{1}^{m+1} R_{2}^{m} (S_{0}^{m})^{2} S_{1}^{m}+(R_{2}^{m})^{2} (R_{2}^{m+1})^{2} (S_{1}^{m})^{2}+2 R_{1}^{m+1} R_{2}^{m} R_{2}^{m+1} S_{0}^{m} (S_{1}^{m})^{2}+2 (R_{2}^{m})^{2} R_{2}^{m+1} S_{0}^{m} (S_{1}^{m})^{2}\nonumber\\
&\fl+(R_{1}^{m+1})^{2} (S_{0}^{m})^{2} (S_{1}^{m})^{2}+2 R_{1}^{m+1} R_{2}^{m} (S_{0}^{m})^{2} (S_{1}^{m})^{2}+(R_{2}^{m})^{2} (S_{0}^{m})^{2} (S_{1}^{m})^{2}+6 R_{0}^{m} R_{0}^{m+1} S_{2}^{m}+2 R_{0}^{m} R_{1}^{m} S_{2}^{m}\nonumber\\
&\fl+2 R_{0}^{m+1} R_{2}^{m+1} S_{2}^{m}+2 R_{1}^{m} R_{2}^{m+1} S_{2}^{m}+6 R_{0}^{m+1} S_{0}^{m} S_{2}^{m}+6 R_{1}^{m} S_{0}^{m} S_{2}^{m}+2 R_{0}^{m} R_{0}^{m+1} R_{1}^{m} R_{1}^{m+1} S_{0}^{m} S_{2}^{m}\nonumber\\
&\fl+2 R_{0}^{m+1} R_{1}^{m} R_{1}^{m+1} (S_{0}^{m})^{2} S_{2}^{m}+2 (R_{1}^{m})^{2} R_{1}^{m+1} (S_{0}^{m})^{2} S_{2}^{m}+6 R_{0}^{m} S_{1}^{m} S_{2}^{m}+6 R_{2}^{m+1} S_{1}^{m} S_{2}^{m}\nonumber\\
&\fl+2 R_{0}^{m} R_{0}^{m+1} R_{2}^{m} R_{2}^{m+1} S_{1}^{m} S_{2}^{m}+6 S_{0}^{m} S_{1}^{m} S_{2}^{m}+2 R_{1}^{m} R_{1}^{m+1} (S_{0}^{m})^{2} S_{1}^{m} S_{2}^{m}+2 R_{0}^{m} R_{2}^{m} R_{2}^{m+1} (S_{1}^{m})^{2} S_{2}^{m}\nonumber\\
&\fl+2 R_{2}^{m} (R_{2}^{m+1})^{2} (S_{1}^{m})^{2} S_{2}^{m}+2 R_{2}^{m} R_{2}^{m+1} S_{0}^{m} (S_{1}^{m})^{2} S_{2}^{m}+(R_{0}^{m})^{2} (R_{0}^{m+1})^{2} (S_{2}^{m})^{2}+2 R_{0}^{m} (R_{0}^{m+1})^{2} S_{0}^{m} (S_{2}^{m})^{2}\nonumber\\
&\fl+2 R_{0}^{m} R_{0}^{m+1} R_{1}^{m} S_{0}^{m} (S_{2}^{m})^{2}+(R_{0}^{m+1})^{2} (S_{0}^{m})^{2} (S_{2}^{m})^{2}+2 R_{0}^{m+1} R_{1}^{m} (S_{0}^{m})^{2} (S_{2}^{m})^{2}+(R_{1}^{m})^{2} (S_{0}^{m})^{2} (S_{2}^{m})^{2}\nonumber\\
&\fl+2 (R_{0}^{m})^{2} R_{0}^{m+1} S_{1}^{m} (S_{2}^{m})^{2}+2 R_{0}^{m} R_{0}^{m+1} R_{2}^{m+1} S_{1}^{m} (S_{2}^{m})^{2}+2 R_{0}^{m} R_{0}^{m+1} S_{0}^{m} S_{1}^{m} (S_{2}^{m})^{2}+(R_{0}^{m})^{2} (S_{1}^{m})^{2} (S_{2}^{m})^{2}\nonumber\\
&\fl+2 R_{0}^{m} R_{2}^{m+1} (S_{1}^{m})^{2} (S_{2}^{m})^{2}+(R_{2}^{m+1})^{2} (S_{1}^{m})^{2} (S_{2}^{m})^{2})/(-1+S_{0}^{m} S_{1}^{m} S_{2}^{m})^2.
\end{eqnarray*}

\end{subsection}

\end{section}

\begin{section}{The Lax pair and conserved quantities of the delay Lotka-Volterra equation}
\label{sec_contiLV}

In this section, we construct the B\"{a}cklund transformation, Lax pair, and conserved quantities of the delay LV equation using the method employed in section \ref{sec_disLV}.

\begin{subsection}{Construction of the delay Lotka-Volterra equation}

First, applying the delay-differential limit
\begin{eqnarray}
    m\delta=t,\qquad\alpha\delta=2\tau,\qquad\delta\to0\label{dlydiffLimitLV}
\end{eqnarray}
to the discrete-time delay LV equation (\ref{dlydisLVb}) and (\ref{dlydisLVn}), we can obtain the bilinear form of the delay LV equation~\cite{Nakata1}:
\begin{eqnarray}
    \fl D_{t}f_{n+\beta}(t+\tau)\cdot f_{n-1}(t-\tau)-f_{n+1+\beta}(t+\tau)f_{n-2}(t-\tau)+f_{n+\beta}(t+\tau)f_{n-1}(t-\tau)=0\label{dlyLVb}
\end{eqnarray}
and the nonlinear form of the delay LV equation 
\begin{eqnarray}
    \fl \frac{d}{dt}\log \frac{u_{n+\beta}(t+\tau)}{u_{n-1}(t-\tau)}=u_{n+1+\beta}(t+\tau)-u_{n+\beta}(t+\tau)-u_{n-1}(t-\tau)+u_{n-2}(t-\tau),\label{dlyLVn}
\end{eqnarray}
where the dependent variable transformation is
\begin{eqnarray}
    u_{n}(t)=\frac{f_{n+1+\beta}(t+2\tau)f_{n-2}(t)}{f_{n+\beta}(t+2\tau)f_{n-1}(t)}.\label{dlyLVtrans}
\end{eqnarray}
Equation (\ref{dlyLVn}) leads to the subtraction of the following LV equations when $\tau=\beta=0$:
\begin{eqnarray}
    \fl \frac{d}{dt}\log u_{n}(t)=u_{n+1}(t)-u_{n-1}(t),\qquad
    \frac{d}{dt}\log u_{n-1}(t)=u_{n}(t)-u_{n-2}(t).\label{LVn}
\end{eqnarray}
By introducing another dependent variable
\begin{equation}
    T_{n}(t)
    =\frac{d}{dt}\log\frac{f_{n}(t)}{f_{n-1}(t)},
    %=\sum_{k=1}^{\infty}\left(u_{n+1-k(1+\beta)}(t-2k\tau)-u_{n-k(1+\beta)}(t-2k\tau)\right),
    \label{dlyLVT}
\end{equation}
equation (\ref{dlyLVb}) is transformed into
\begin{eqnarray}
    \eqalign{
    \frac{d}{dt}\log u_{n}(t)=T_{n+1+\beta}(t+2\tau)-T_{n-1}(t),\\
    T_{n+1+\beta}(t+2\tau)-T_{n}(t)=u_{n+1}(t)-u_{n}(t).}\label{dlyLVn_2var}
\end{eqnarray}

\end{subsection}

\begin{subsection}{A B\"{a}cklund transformation and Lax pair of the delay Lotka-Volterra equation}

Next, we construct a B\"{a}cklund transformation of the delay LV equation by applying the delay-differential limit to that of the discrete-time delay LV equation.
Let us replace $\lambda_{1}$ by $\lambda_{1}+1/\delta$ and set $\mu=1/\delta$ in equations (\ref{dlydisLVBT1}) and (\ref{dlydisLVBT2}), and then apply the delay-differential limit (\ref{dlydiffLimitLV}).
Then a B\"{a}cklund transformation of the delay LV equation is obtained as follows:
\begin{eqnarray}
    \fl\lambda_{1}f_{n}(t)g_{n+1}(t)+D_{t}f_{n}(t)\cdot g_{n+1}(t)-f_{n+1}(t)g_{n}(t)=0,\label{dlyLVBT1}\\
    \fl\lambda_{2}f_{n+1+\beta}(t+\tau)g_{n}(t-\tau)+f_{n}(t-\tau)g_{n+1+\beta}(t+\tau)-f_{n-1}(t-\tau)g_{n+2+\beta}(t+\tau)=0,\label{dlyLVBT2}
\end{eqnarray}
where $g$ is a new solution of equation (\ref{dlyLVb}) obtained from the solution $f$.

Now, we construct the Lax pair of the delay LV equation.
By putting $g_{n}(t)=f_{n}(t)\psi_{n}(t)$, equations (\ref{dlyLVBT1}) and (\ref{dlyLVBT2}) lead to
\begin{eqnarray}
    \frac{d}{dt}\psi_{n}(t)+T_{n}(t)\psi_{n}(t)+\psi_{n-1}(t)=\lambda_{1}\psi_{n}(t),\label{dlyLVLax1}\\
    u_{n+1}(t)\psi_{n+2+\beta}(t+2\tau)-\psi_{n+1+\beta}(t+2\tau)=\lambda_{2}\psi_{n}(t),\label{dlyLVLax2}
\end{eqnarray}
where $u_{n}(t)$ and $T_{n}(t)$ are defined by (\ref{dlyLVtrans}) and (\ref{dlyLVT}).
We introduce linear operators $M$ and $L_{2}$ by
\begin{eqnarray}
    M(n,t)=T_{n}(t)+\exp\left(-\frac{\partial}{\partial n}\right),\\
    L_{2}(n,t)=u_{n+1}(t)\exp\left((2+\beta)\frac{\partial}{\partial n}\right)
    -\exp\left((1+\beta)\frac{\partial}{\partial n}\right).
\end{eqnarray}
Then equations (\ref{dlyLVLax1}) and (\ref{dlyLVLax2}) are rewritten as follows:
\begin{eqnarray}
    \frac{d}{dt}\psi_{n}(t)+M(n,t)\psi_{n}(t)
    =\lambda_{1}\psi_{n}(t),\label{dlyLVLaxM}\\
    L_{2}(n,t)\psi_{n}(t+2\tau)\label{dlyLVLaxL2}
    =\lambda_{2}\psi_{n}(t).
\end{eqnarray}
The compatibility condition of equations (\ref{dlyLVLaxM}) and (\ref{dlyLVLaxL2}) is 
\begin{eqnarray}
    \frac{d}{dt}L_{2}(n,t)=L_{2}(n,t)M(n,t+2\tau)-M(n,t)L_{2}(n,t).\label{dlyLVcompa}
\end{eqnarray}
Equation (\ref{dlyLVcompa}) leads to the delay LV equation (\ref{dlyLVn}) or (\ref{dlyLVn_2var}).
\begin{rem}
When $\beta=\tau=0$, equations (\ref{dlyLVLax1}) and (\ref{dlyLVLax2}) lead to the following equations:
\begin{eqnarray}
    \frac{d}{dt}\psi_{n}(t)+\left(-1+u_{n}(t)\right)\psi_{n}(t)+\psi_{n-1}(t)=\lambda_{1}\psi_{n}(t),\label{LVLax1}\\
    u_{n+1}(t)\psi_{n+2}(t)-\psi_{n+1}(t)=\lambda_{2}\psi_{n}(t).\label{LVLax}
\end{eqnarray}
The compatibility condition of these equations leads to the LV equation (\ref{LVn}).
\end{rem}
% \begin{rem}
% Note that we can also obtain the relation
% \begin{eqnarray}
% \frac{d}{dt}L_{2}(n,t)^{-1}=L_{2}(n,t)^{-1}M(n,t)-M(n,t+2\tau)L_{2}(n,t)^{-1}
% \end{eqnarray}
% by using equation (\ref{dlyLVcompa}) and the property
% \begin{eqnarray}
% 0=\frac{d}{dt}\left(A(t)A(t)^{-1}\right)=\frac{d}{dt}\left(A(t)\right)A(t)^{-1}+A(t)\frac{d}{dt}\left(A(t)^{-1}\right).
% \end{eqnarray}
% for a matrix $A(t)$.
% \end{rem}

\end{subsection}

\begin{subsection}{The Lax pair and conserved quantities of the delay Lotka-Volterra equation under the periodic boundary condition}

Now, we introduce the $N$-periodic boundary condition $\psi_{n+N}(t)=\psi_{n}(t)$ in the space variable $n$.
By defining 
\begin{eqnarray}
    \hat{\psi}(t)=(\psi_{0}(t),\psi_{1}(t),\psi_{2}(t),\cdots,\psi_{N-1}(t))^{T},
\end{eqnarray}
equations (\ref{dlyLVLaxM}) and (\ref{dlyLVLaxL2}) in the case of $n=0,1,\cdots,N-1$ can be expressed in the following form:
\begin{eqnarray}
    \frac{d}{dt}\hat{\psi}(t)+\mathcal{M}(t)\hat{\psi}(t)
    =\lambda_{1}\hat{\psi}(t),\label{dlyLVLaxMatrix1}\\
    \mathcal{L}_{2}(t)\hat{\psi}(t+2\tau)
    =\lambda_{2}\hat{\psi}(t),\label{dlyLVLaxMatrix2}
\end{eqnarray}
where $\mathcal{M}(t)$ and $\mathcal{L}_{2}(t)$ are described by matrices.
For example, they are defined as follows in the case of $\beta=2$:
\begin{eqnarray}
\fl \mathcal{M}(t)=
\left(
\begin{array}{ccccc}
T_{0}(t) & 0        & 0        & \cdots & 1 \\
1        & T_{1}(t) & 0        & \cdots & 0 \\
0        & 1        & T_{2}(t) & \cdots & 0 \\
\vdots   & \vdots   & \ddots   & \ddots & \vdots \\
0        & 0        & \cdots   & 1      & T_{N-1}(t) \\
\end{array}
\right),\label{dlyLVmatrixM}\\
\fl \mathcal{L}_{2}(t)=
\left(
\begin{array}{cccccccc}
0      & 0      & 0      & -1     & u_{1}(t) & 0        & \cdots & 0\\
0      & 0      & 0      & 0      & -1       & u_{2}(t) & \cdots & 0\\
\vdots & \vdots & \vdots & \vdots & \ddots   & \ddots   & \ddots & \vdots\\
0      & 0      & 0      & 0      & \cdots   & 0        & -1     & u_{N-4}(t)\\
u_{N-3}(t) & 0 & 0 & 0            & \cdots   & 0        & 0      & -1\\
-1 & u_{N-2}(t) & 0 & 0           & \cdots   & 0        & 0      & 0\\
0 & -1 & u_{N-1}(t) & 0           & \cdots   & 0        & 0      & 0\\
0 & 0 & -1 & u_{0}(t)             & \cdots   & 0        & 0      & 0\\
\end{array}
\right).\label{dlyLVmatrixL2}
\end{eqnarray}

Now, by using the above Lax pair, we construct conserved quantities as functions of $t$ under the $N$-periodic boundary condition in the spatial variable $n$.
The compatibility condition of equations (\ref{dlyLVLaxMatrix1}) and (\ref{dlyLVLaxMatrix2}) is
\begin{eqnarray}
    \frac{d}{dt}\mathcal{L}_{2}(n,t)
    =\mathcal{L}_{2}(n,t)\mathcal{M}(n,t+2\tau)-\mathcal{M}(n,t)\mathcal{L}_{2}(n,t).\label{dlyLVMatrix_compa1}
\end{eqnarray}
In the case of $\tau=0$, the function $\mathrm{Tr}\left(\mathcal{L}_{2}(m)^{k}\right)$ are conserved quantities.
On the other hand, when $\tau\neq0$, equation (\ref{dlyLVMatrix_compa1}) cannot be used directly to construct conserved quantities, because the delay $\tau$ is contained and the right side of the equation cannot be expressed in terms of commutators.
It is necessary to apply the delay-differential limit to equation (\ref{dlydisLVMatrix_compa2}).

To consider the delay-differential limit of (\ref{dlydisLVMatrix_compa2}), we impose $\alpha\geq1$ and define a new discrete linear operator 
\begin{eqnarray}
    \bar{\mathcal{N}}^{*}(m)&=&\delta^{\alpha}\bar{\mathcal{N}}(m)\nonumber\\
    &=&\delta^{\alpha}\bar{\mathcal{L}}_{1}(m)\bar{\mathcal{L}}_{1}(m+1)\cdots \bar{\mathcal{L}}_{1}(m+\alpha-1)\bar{\mathcal{L}}_{2}(m)^{-1}.\label{dlydisLVop1}
\end{eqnarray}
% where $\bar{\mathcal{L}}_{1}(m)$ and $\bar{\mathcal{L}}_{2}(m)$ are defined by (\ref{dlydisLVLaxMatrixL1}) and (\ref{dlydisLVLaxMatrixL2}).
% Equation (\ref{dlydisLVop1}) can be rewritten by
This can be rewritten by
\begin{eqnarray}
    \fl \bar{\mathcal{N}}^{*}(m)=(I+\delta \bar{\mathcal{M}}(m))(I+\delta \bar{\mathcal{M}}(m+1))\cdots(I+\delta \bar{\mathcal{M}}(m+\alpha-1))\bar{\mathcal{L}}_{2}(m)^{-1},\label{dlydisLVop2}
\end{eqnarray}
where $\bar{\mathcal{M}}(m)$ is defined by
\begin{eqnarray}
    \bar{\mathcal{L}}_{1}(m)=\frac{1}{\delta}I+\bar{\mathcal{M}}(m)\label{dlydisLVMbar}
\end{eqnarray} 
and $I$ is the identity matrix.
Since $\mu=1/\delta$, we can show that $\bar{\mathcal{M}}(m)$ and $\bar{\mathcal{L}}_{2}(m)$ are reduced to $\mathcal{M}(t)$ and $\mathcal{L}_{2}(t)$ as $\delta\to0$ respectively.
Applying the delay-differential limit (\ref{dlydiffLimitLV}) to (\ref{dlydisLVop2}), we obtain the following $\mathcal{N}^{*}(t)$ as the limit of $\bar{\mathcal{N}}^{*}(m)$:
\begin{eqnarray}
    \mathcal{N}^{*}(t)=E(t)\mathcal{L}_{2}(t)^{-1},\label{dlyLVop2}
\end{eqnarray}
where
\begin{eqnarray}
    \fl E(t)=&I+\int_{0}^{2\tau}dx_{1} \mathcal{M}(t+x_{1})+\int_{0}^{2\tau}dx_{2}\int_{0}^{x_{2}}dx_{1} \mathcal{M}(t+x_{1})\mathcal{M}(t+x_{2})\nonumber\\
    \fl &+\int_{0}^{2\tau}dx_{3}\int_{0}^{x_{3}}dx_{2}\int_{0}^{x_{2}}dx_{1} \mathcal{M}(t+x_{1})\mathcal{M}(t+x_{2})\mathcal{M}(t+x_{3})+\cdots.\label{dlyLV_E_2}
\end{eqnarray}
% Using (\ref{dlyLVcompa}) and
% \begin{eqnarray}
%    \frac{d}{dt}E(t)=E(t)M(t+2\tau)-\mathcal{M}(t)E(t)\,,
% \end{eqnarray}
The details of these calculations are written in \ref{sec_derivation}.
On the other hand, as the delay-differential limit (\ref{dlydiffLimitLV}) of equation (\ref{dlydisLVMatrix_compa2}), we obtain
\begin{eqnarray}
    \frac{d}{dt}\mathcal{N}^{*}(t)=\mathcal{N}^{*}(t)\mathcal{M}(t)-\mathcal{M}(t)\mathcal{N}^{*}(t).\label{dlyLVMatrix_compa2}
\end{eqnarray}
Therefore, defining $H_{k}(t)=\mathrm{Tr}\left(\mathcal{N}^{*}(t)^{k}\right)$, we obtain
\begin{eqnarray}
    \frac{d}{dt}H_{k}(t)
    =\mathrm{Tr}\left(k\mathcal{N}^{*}(t)^{k-1}\frac{d}{dt}\mathcal{N}^{*}(t)\right)
    =0.
\end{eqnarray}
Thus $H_{k}(t)\ (k=1,2,\cdots)$ are conserved quantities of the delay LV equation.

We remark that $E(t)$, which is a part of the conserved quantities $H_{k}(t)$, is rewritten by 
\begin{eqnarray}
    E(t)=T\left(\exp\left(\int_{0}^{2\tau}\mathcal{M}(t+s)ds\right)\right).\label{dlyLV_E}
\end{eqnarray}
$T$ is the time-ordered product defined by 
\begin{eqnarray}
    T\left(A(t_{1})A(t_{2})\cdots A(t_{i})\right)=A(t_{\sigma(1)})A(t_{\sigma(2)})\cdots A(t_{\sigma(i)}),\label{time-ordered_product}
\end{eqnarray}
where $\sigma$ is a permutation that satisfies $t_{\sigma(1)}\leq t_{\sigma(2)}\leq\cdots\leq t_{\sigma(i)}$.
$T$ sorts $A(t_1),\cdots,A(t_i)$ in increasing order of time.
The details of the calculation are written in \ref{sec_derivation}.
The time-ordered product appears in, for example, quantum field theory~\cite{peskin1}.

The fact that $\bar{\mathcal{L}}_{1}$ is acted on $\alpha$ times in (\ref{dlydisLVop1}) and the delay-differential limit (\ref{dlydiffLimitLV}), which includes $\left|\alpha\right|\to\infty$, cause the exponential structure (\ref{dlyLV_E}) in the conserved quantities $H_{k}(t)$.
This is an effect caused by the delay.

\end{subsection}

\end{section}

\begin{section}{The Lax pair and conserved quantities of the discrete-time delay Toda lattice equation}
\label{sec_disTL}

In this section, we construct the B\"{a}cklund transformation, Lax pair, and conserved quantities of the discrete-time delay Toda lattice equation, which can be considered as a higher order discrete TL equation, before considering the delay Toda lattice equation.

\begin{subsection}{Construction of the discrete-time delay Toda lattice equation}

We first consider a reduction of the DAGTE (\ref{DAGTE}).
By setting
\begin{eqnarray}
    \eqalign{
    Z_{1}=c\delta,\quad Z_{2}=1-c\delta,\quad Z_{3}=-1,\\
    D_{1}=\frac{\alpha D_{m}+(2+\beta)D_n}{2},\quad
    D_{2}=-\frac{\alpha D_{m}+\beta D_n}{2},\\
    D_{3}=-\frac{(2+\alpha)D_{m}+\beta D_n}{2},}\label{reductionTL}
\end{eqnarray}
equation (\ref{DAGTE}) is reduced to the bilinear form of the discrete-time delay Toda lattice equation~\cite{Nakata1}:
\begin{eqnarray}
    f_{n+\beta}^{m+1+\alpha}f_{n}^{m-1}
    -c\delta f_{n+1+\beta}^{m+\alpha}f_{n-1}^{m}
    -(1-c\delta)f_{n+\beta}^{m+\alpha}f_{n}^{m}=0,
    \label{dlydisTLb}
\end{eqnarray}
where $\alpha$ and $\beta$ are constant integers and $c$ is a real constant.
The nonlinear form of the discrete-time delay Toda lattice equation 
\begin{eqnarray}
   \frac{(1+V_{n+\beta}^{m+1+\alpha})(1+V_{n}^{m-1})}{(1+V_{n+\beta}^{m+\alpha})(1+V_{n}^{m})}=\frac{(1+c\delta V_{n+1+\beta}^{m+\alpha})(1+c\delta V_{n-1}^{m})}{(1+c\delta V_{n+\beta}^{m+\alpha})(1+c\delta V_{n}^{m})}\label{dlydisTLn}
\end{eqnarray}
is obtained through the dependent variable transformation
\begin{eqnarray}
   1+V_{n}^{m}=\frac{f_{n+1+\beta}^{m+\alpha}f_{n-1}^{m}}{f_{n+\beta}^{m+\alpha}f_{n}^{m}}.\label{dlydisTLtrans}
\end{eqnarray}
Equation (\ref{dlydisTLn}) leads to the following discrete Toda lattice equation when $\alpha=\beta=0,\ c=\delta$:
\begin{eqnarray}
    \frac{(1+V_{n}^{m+1})(1+V_{n}^{m-1})}{(1+V_{n}^{m})(1+V_{n}^{m})}=\frac{(1+\delta^2 V_{n+1}^{m})(1+\delta^2 V_{n-1}^{m})}{(1+\delta^2 V_{n}^{m})(1+\delta^2 V_{n}^{m})}.\label{disTLn}
\end{eqnarray}
By introducing the alternative dependent variables
\begin{eqnarray}
    I_{n}^{m}
    =\frac{f_{n}^{m-1}f_{n+1}^{m}}{f_{n+1}^{m-1}f_{n}^{m}},\qquad
    %=\prod_{k=1}^{\infty}\frac{1+c\delta V_{n+1-k\beta}^{m-k(1+\alpha)}}{1+c\delta V_{n-k\beta}^{m-k(1+\alpha)}},\\
    J_{n}^{m}
    =\frac{c\delta}{1-c\delta}\frac{f_{n-1-\beta}^{m-1-\alpha}f_{n+1}^{m}}{f_{n-\beta}^{m-1-\alpha}f_{n}^{m}},
    %=\frac{c\delta(1+V_{n-\beta}^{m-1-\alpha})}{1-c\delta}\prod_{k=1}^{\infty}\frac{1+c\delta V_{n+1-k\beta}^{m-k(1+\alpha)}}{1+c\delta V_{n-k\beta}^{m-k(1+\alpha)}},
    \label{dlydisTLIJ}
\end{eqnarray}
equation (\ref{dlydisTLb}) is transformed into
\begin{eqnarray}
    \eqalign{
    I_{n}^{m}J_{n}^{m-1}=I_{n-1-\beta}^{m-1-\alpha}J_{n}^{m},\\
    I_{n+\beta}^{m+1+\alpha}-I_{n}^{m}=J_{n+1+\beta}^{m+\alpha}-J_{n+\beta}^{m+1+\alpha}.}\label{dlydisTLn_2var}
\end{eqnarray}

\end{subsection}

\begin{subsection}{A B\"{a}cklund transformation and Lax pair of the discrete-time delay Toda lattice equation}

By applying the above reduction condition (\ref{reductionTL}) to equations (\ref{DAGTEBT1}) and (\ref{DAGTEBT2}), a B\"{a}cklund transformation of the discrete-time delay Toda lattice equation (\ref{dlydisTLb}) can be obtained as follows:
\begin{eqnarray}
    \lambda_{1}f_{n+1}^{m}g_{n}^{m+1}-f_{n}^{m+1}g_{n+1}^{m}-\mu f_{n+1}^{m+1}g_{n}^{m}=0,\label{dlydisTLBT1}\\
    \lambda_{2}f_{n}^{m}g_{n+\beta}^{m+1+\alpha}+f_{n+\beta}^{m+1+\alpha}g_{n}^{m}-\frac{c\mu\delta}{1-c\delta}f_{n+1+\beta}^{m+1+\alpha}g_{n-1}^{m}=0,\label{dlydisTLBT2}
\end{eqnarray}
where
%$\mu_{1}=\mu,\ \mu_{2}=-c\mu\delta/(1-c\delta)$ and
$\mu$ is an arbitrary constant.

Next, we construct the Lax pair of the discrete-time delay Toda lattice equation.
Setting $g_{n}^{m}=f_{n}^{m}\psi_{n}^{m}$, we obtain the following equations by (\ref{dlydisTLBT1}) and (\ref{dlydisTLBT2}):
\begin{eqnarray}
    \mu I_{n}^{m}\psi_{n}^{m-1}
    +\psi_{n+1}^{m-1}
    =\lambda_{1} \psi_{n}^{m},\label{dlydisTLLax1}\\
    \mu J_{n}^{m}\psi_{n-1-\beta}^{m-1-\alpha}
    -\psi_{n-\beta}^{m-1-\alpha}
    =\lambda_{2}\psi_{n}^{m},\label{dlydisTLLax2}
\end{eqnarray}
where $I_{n}^{m}$ and $J_{n}^{m}$ are defined by (\ref{dlydisTLIJ}).
We introduce linear operators $\bar{L_{1}}$ and $\bar{L_{2}}$ by
\begin{eqnarray}
    \bar{L_{1}}(n,m)=\mu I_{n}^{m}+\exp\left(\frac{\partial}{\partial n}\right),\\
    \bar{L_{2}}(n,m)=\mu J_{n}^{m}\exp\left(-(1+\beta)\frac{\partial}{\partial n}\right)-\exp\left(-\beta\frac{\partial}{\partial n}\right).
\end{eqnarray}
Then equations (\ref{dlydisTLLax1}) and (\ref{dlydisTLLax2}) are rewritten as follows:
\begin{eqnarray}
    \bar{L_{1}}(n,m)\psi_{n}^{m-1}=\lambda_{1}\psi_{n}^{m},\label{dlydisTLLaxL1}\\
    \bar{L_{2}}(n,m)\psi_{n}^{m-1-\alpha}=\lambda_{2}\psi_{n}^{m}.\label{dlydisTLLaxL2}
\end{eqnarray}
The compatibility condition of equations (\ref{dlydisTLLaxL1}) and (\ref{dlydisTLLaxL2}) is
\begin{eqnarray}
    \bar{L_{1}}(n,m)\bar{L_{2}}(n,m-1)
    =\bar{L_{2}}(n,m)\bar{L_{1}}(n,m-1-\alpha),\label{dlydisTLcompa}
\end{eqnarray}
and this equation leads to the discrete-time delay Toda lattice equation (\ref{dlydisTLn}) or (\ref{dlydisTLn_2var}).

\end{subsection}

\begin{subsection}{The Lax pair and conserved quantities of the discrete-time delay Toda lattice equation under the periodic boundary condition}

Now, we introduce the $N$-periodic boundary condition  $\psi_{n+N}^{m}=\psi_{n}^{m}$ in the space variable $n$.
By defining 
\begin{eqnarray}
    \hat{\psi}(m)=(\psi_{0}^{m},\psi_{1}^{m},\psi_{2}^{m},\cdots,\psi_{N-1}^{m})^{T},
\end{eqnarray}
equations (\ref{dlydisTLLaxL1}) and (\ref{dlydisTLLaxL2}) in the case of $n=0,1,\cdots,N-1$ can be expressed in the following form:
\begin{eqnarray}
    \bar{\mathcal{L}}_{1}(m)\hat{\psi}(m-1)=\lambda_{1}\hat{\psi}(m),\label{dlydisTLLaxMatrix1}\\
    \bar{\mathcal{L}}_{2}(m)\hat{\psi}(m-1-\alpha)=\lambda_{2}\hat{\psi}(m)\label{dlydisTLLaxMatrix2},
\end{eqnarray}
where $\bar{\mathcal{L}}_{1}(m)$ and $\bar{\mathcal{L}}_{2}(m)$ are described by matrices.
For example, they are defined as follows in the case of $\beta=0$:
\begin{eqnarray}
\bar{\mathcal{L}}_{1}(m)=
\left(
\begin{array}{ccccc}
\mu I_{0}^{m} & 1             & 0             & \cdots & 0 \\
0             & \mu I_{1}^{m} & 1             & \cdots & 0 \\
0             & 0             & \mu I_{2}^{m} & \cdots & 0 \\
\vdots        & \vdots        & \ddots        & \ddots & \vdots \\
1             & 0             & \cdots        &0       & \mu I_{N-1}^{m} \\
\end{array}
\right),\label{dlydisTLLaxMatrixL1}\\
\bar{\mathcal{L}}_{2}(m)=
\left(
\begin{array}{ccccc}
-1            & 0             & 0         & \cdots & \mu J_{0}^{m} \\
\mu J_{1}^{m} & -1            & 0         & \cdots & 0 \\
0             & \mu J_{2}^{m} & -1        & \cdots & 0 \\
\vdots        & \vdots        & \ddots    & \ddots & \vdots \\
0             & 0             & \cdots    & \mu J_{N-1}^{m} & -1 \\
\end{array}
\right).\label{dlydisTLLaxMatrixL2}
\end{eqnarray}

Now, by using the above Lax pair, we construct conserved quantities as functions of $m$ under the $N$-periodic boundary condition in the spatial variable $n$.
We consider the case $\alpha\neq-1$ and define a new linear operator 
\begin{eqnarray}
\fl\bar{\mathcal{N}}(m)=
\left\{
\begin{array}{ll}
    \bar{\mathcal{L}}_{1}(m)\bar{\mathcal{L}}_{1}(m-1)\cdots \bar{\mathcal{L}}_{1}(m-\alpha)\bar{\mathcal{L}}_{2}(m)^{-1} & (\alpha\geq0) \\
    \bar{\mathcal{L}}_{2}(m)\bar{\mathcal{L}}_{1}(m-\alpha-1)\bar{\mathcal{L}}_{1}(m-\alpha-2)\cdots \bar{\mathcal{L}}_{1}(m+1) & (\alpha\leq-2)
\end{array}
\right..
\label{dlydisTLop0}
\end{eqnarray}
The compatibility condition of equations (\ref{dlydisTLLaxMatrix1}) and (\ref{dlydisTLLaxMatrix2}) is
\begin{eqnarray}
    \bar{\mathcal{L}}_{1}(m)\bar{\mathcal{L}}_{2}(m-1)
    =\bar{\mathcal{L}}_{2}(m)\bar{\mathcal{L}}_{1}(m-\alpha-1).\label{dlydisTLMatrix_compa1}
\end{eqnarray}
From this condition, the following equation is obtained:
\begin{eqnarray}
    \bar{\mathcal{N}}(m)\bar{\mathcal{L}}_{1}(m)
    =\bar{\mathcal{L}}_{1}(m)\bar{\mathcal{N}}(m-1).\label{dlydisTLMatrix_compa2}
\end{eqnarray}
Defining $\bar{H}_{k}(m)=\mathrm{Tr}\left(\bar{\mathcal{N}}(m)^{k}\right)$, we obtain
\begin{eqnarray}
    \bar{H}_{k}(m+1)
    &=&\mathrm{Tr}\left(\bar{\mathcal{N}}(m+1)^{k}\right)\nonumber\\
    &=&\mathrm{Tr}\left(\bar{\mathcal{L}}_{1}(m+1)\bar{\mathcal{N}}(m)^{k}\bar{\mathcal{L}}_{1}(m+1)^{-1}\right)\nonumber\\
    &=&\mathrm{Tr}\left(\bar{\mathcal{N}}(m)^{k}\right)\nonumber\\
    &=&\bar{H}_{k}(m).
\end{eqnarray}
Thus $\bar{H}_{k}(m)\ (k=1,2,\cdots)$ are conserved quantities of the discrete-time delay Toda lattice equation.
In the case of $\alpha=-1$, we just need to define $\bar{H}_{k}(m):=\mathrm{Tr}\left(\bar{\mathcal{L}_{2}}(m)^{k}\right)$ according to the compatibility condition (\ref{dlydisTLMatrix_compa1}).
% On the other hand, when $\alpha\neq-1$, equation (\ref{dlydisTLMatrix_compa1}) cannot be used directly to construct conserved quantities.
% It is necessary to introduce the operator $\bar{\mathcal{N}}(m)$ to eliminate $\alpha$ in equation (\ref{dlydisTLMatrix_compa1}).
% This is one of the problems caused by the delay.

Let us show examples of $\bar{H}_{k}(m)\ (k=1,2,\cdots)$.
Considering the case of $N=3$ and setting $\alpha=1,\ \beta=0,\ \mu=1$, we can calculate $\bar{H}_{1}(m)$ and $\bar{H}_{2}(m)$ as follows:
\begin{eqnarray*}
\fl\bar{H}_{1}(m)=(I_{0}^{m-1}I_{0}^{m}+I_{1}^{m-1}I_{1}^{m}+I_{2}^{m-1}I_{2}^{m}-I_{0}^{m-1}J_{0}^{m}-I_{1}^{m-1}J_{1}^{m}-I_{2}^{m-1}J_{2}^{m}-I_{0}^{m}J_{1}^{m}\\
-I_{1}^{m}J_{2}^{m}-I_{2}^{m}J_{0}^{m}+J_{0}^{m}J_{1}^{m}+J_{1}^{m}J_{2}^{m}+J_{2}^{m}J_{0}^{m}))/(1+J_{0}^{m}J_{1}^{m}J_{2}^{m}).
\end{eqnarray*}
\begin{eqnarray*}
&\fl \bar{H}_{2}(m)\\
&\fl =((I_{2}^{m-1}I_{2}^{m}-J_{0}^{m}(I_{0}^{m-1}+I_{2}^{m}-J_{1}^{m}))^2+(I_{0}^{m-1} I_{0}^{m}-J_{1}^{m} (I_{0}^{m}+I_{1}^{m-1}-J_{2}^{m}))^2\nonumber\\
&\fl+(I_{1}^{m-1} I_{1}^{m}-(I_{1}^{m}+I_{2}^{m-1}-J_{0}^{m}) J_{2}^{m})^2\nonumber\\
&\fl+2 (I_{2}^{m-1}-J_{0}^{m}+I_{1}^{m} (1+I_{1}^{m-1} J_{0}^{m} J_{1}^{m})) (1+(-I_{2}^{m-1} I_{2}^{m}+(I_{0}^{m-1}+I_{2}^{m}) J_{0}^{m}) J_{2}^{m})\nonumber\\
&\fl+2 (I_{0}^{m}+I_{1}^{m-1}+(-1+I_{0}^{m-1} I_{0}^{m} J_{0}^{m}) J_{2}^{m}) (1+J_{1}^{m} (-I_{1}^{m-1} I_{1}^{m}+(I_{1}^{m}+I_{2}^{m-1}) J_{2}^{m}))\nonumber\\
&\fl+2 (1+J_{0}^{m} (-I_{0}^{m-1} I_{0}^{m}+(I_{0}^{m}+I_{1}^{m-1}) J_{1}^{m})) (I_{0}^{m-1}-J_{1}^{m}+I_{2}^{m} (1+I_{2}^{m-1} J_{1}^{m} J_{2}^{m})))/(1+J_{0}^{m} J_{1}^{m} J_{2}^{m})^2.
\end{eqnarray*}

\end{subsection}

\end{section}

\begin{section}{The Lax pair and conserved quantities of the delay Toda lattice equation}
\label{sec_contiTL}

In this section, we construct the B\"{a}cklund transformation, Lax pair, and conserved quantities of the delay Toda lattice equation using the method employed in section \ref{sec_disTL}.

\begin{subsection}{Construction of the delay Toda lattice equation}

First, applying the delay-differential limit
\begin{eqnarray}
    m\delta=t,\qquad c=\alpha\delta=2\tau,\qquad\delta\to0\label{dlydiffLimitTL}
\end{eqnarray}
to the discrete-time delay Toda lattice equation (\ref{dlydisTLb}) and (\ref{dlydisTLn}), we can obtain the bilinear form of the delay Toda lattice equation~\cite{Nakata1}:
\begin{eqnarray}
    \eqalign{
    \fl D_{t}f_{n+\beta}(t+\tau)\cdot f_{n}(t-\tau)\\
    -2\tau(f_{n+1+\beta}(t+\tau)f_{n-1}(t-\tau)-f_{n+\beta}(t+\tau)f_{n}(t-\tau))=0}\label{dlyTLb}
\end{eqnarray}
and the nonlinear form of the delay Toda lattice equation 
\begin{eqnarray}
    \fl \frac{1}{2\tau}\frac{d}{dt}\log \frac{1+V_{n+\beta}(t+\tau)}{1+V_{n}(t-\tau)}=V_{n+1+\beta}(t+\tau)-V_{n+\beta}(t+\tau)-V_{n}(t-\tau)+V_{n-1}(t-\tau),\label{dlyTLn}
\end{eqnarray}
where the dependent variable transformation is
\begin{eqnarray}
    1+V_{n}(t)=\frac{f_{n+1+\beta}(t+2\tau)f_{n-1}(t)}{f_{n+\beta}(t+2\tau)f_{n}(t)}.\label{dlyTLtrans}
\end{eqnarray}
Equation (\ref{dlyTLn}) leads to the following Toda lattice equation when $\tau\to0$ and $\beta=0$:
\begin{eqnarray}
    \frac{d^{2}}{dt^{2}}\log (1+V_{n}(t))=V_{n+1}(t)-2V_{n}(t)+V_{n-1}(t).\label{TLn}
\end{eqnarray}
By introducing another dependent variable
\begin{equation}
    T_{n}(t)
    =\frac{d}{dt}\log\frac{f_{n+1}(t)}{f_{n}(t)},
    %=2\tau\sum_{k=1}^{\infty}\left(V_{n+1-k\beta}(t-2k\tau)-V_{n-k\beta}(t-2k\tau)\right),
    \label{dlyTLT}
\end{equation}
equation (\ref{dlyTLb}) is transformed into
\begin{eqnarray}
    \eqalign{
    \frac{d}{dt}\log(1+V_{n}(t))=T_{n+\beta}(t+2\tau)-T_{n-1}(t),\\
    T_{n+\beta}(t+2\tau)-T_{n}(t)=2\tau(V_{n+1}(t)-V_{n}(t)).}\label{dlyTLn_2var}
\end{eqnarray}

\end{subsection}

\begin{subsection}{A B\"{a}cklund transformation and Lax pair of the delay Toda lattice equation}

Next, we construct a B\"{a}cklund transformation of the delay Toda lattice equation by applying the delay-differential limit to that of the discrete-time delay Toda lattice equation.
Let us replace $\lambda_{1}$ by $\lambda_{1}+1/\delta$ and set $\mu=1/\delta$ in equations (\ref{dlydisTLBT1}) and (\ref{dlydisTLBT2}), and then apply the delay-differential limit (\ref{dlydiffLimitTL}).
Then a B\"{a}cklund transformation of the delay Toda lattice equation is obtained as follows:
\begin{eqnarray}
    \fl\lambda_{1}f_{n+1}(t)g_{n}(t)-D_{t}f_{n+1}(t)\cdot g_{n}(t)-f_{n}(t)g_{n+1}(t)=0,\label{dlyTLBT1}\\
    \fl\lambda_{2}f_{n}(t-\tau)g_{n+\beta}(t+\tau)
    +f_{n+\beta}(t+\tau)g_{n}(t-\tau)
    -2\tau f_{n+1+\beta}(t+\tau)g_{n-1}(t-\tau)=0,\label{dlyTLBT2}
\end{eqnarray}
where $g$ is a new solution of equation (\ref{dlyTLb}) obtained from the solution $f$.

Now, we construct the Lax pair of the delay Toda lattice equation.
By putting $g_{n}(t)=f_{n}(t)\psi_{n}(t)$, equations (\ref{dlyTLBT1}) and (\ref{dlyTLBT2}) lead to
\begin{eqnarray}
    -\frac{d}{dt}\psi_{n}(t)+T_{n}(t)\psi_{n}(t)+\psi_{n+1}(t)=\lambda_{1}\psi_{n}(t),\label{dlyTLLax1}\\
    2\tau(1+V_{n-\beta}(t-2\tau))\psi_{n-1-\beta}(t-2\tau)-\psi_{n-\beta}(t-2\tau)=\lambda_{2}\psi_{n}(t),\label{dlyTLLax2}
\end{eqnarray}
where $V_{n}(t)$ and $T_{n}(t)$ are defined by (\ref{dlyTLtrans}) and (\ref{dlyTLT}).
We introduce linear operators $M$ and $L_{2}$ by
\begin{eqnarray}
    \fl M(n,t)=T_{n}(t)+\exp\left(\frac{\partial}{\partial n}\right),\\
    \fl L_{2}(n,t)=2\tau(1+V_{n-\beta}(t-2\tau))\exp\left(-(1+\beta)\frac{\partial}{\partial n}\right)-\exp\left(-\beta\frac{\partial}{\partial n}\right).
\end{eqnarray}
Then equations (\ref{dlyTLLax1}) and (\ref{dlyTLLax2}) are rewritten as follows:
\begin{eqnarray}
    -\frac{d}{dt}\psi_{n}(t)+M(n,t)\psi_{n}(t)
    =\lambda_{1}\psi_{n}(t),\label{dlyTLLaxM}\\
    L_{2}(n,t)\psi_{n}(t-2\tau)\label{dlyTLLaxL2}
    =\lambda_{2}\psi_{n}(t).
\end{eqnarray}
The compatibility condition of equations (\ref{dlyTLLaxM}) and (\ref{dlyTLLaxL2}) is 
\begin{eqnarray}
    \frac{d}{dt}L_{2}(n,t)=M(n,t)L_{2}(n,t)-L_{2}(n,t)M(n,t-2\tau).\label{dlyTLcompa}
\end{eqnarray}
Equation (\ref{dlyTLcompa}) leads to the delay Toda lattice equation (\ref{dlyTLn}) or (\ref{dlyTLn_2var}).
% \begin{rem}
% Note that we can also obtain the relation
% \begin{eqnarray}
% \frac{d}{dt}L_{2}(n,t)^{-1}=L_{2}(n,t)^{-1}M(n,t)-M(n,t+2\tau)L_{2}(n,t)^{-1}
% \end{eqnarray}
% by using equation (\ref{dlyTLcompa}) and the property
% \begin{eqnarray}
% 0=\frac{d}{dt}\left(A(t)A(t)^{-1}\right)=\frac{d}{dt}\left(A(t)\right)A(t)^{-1}+A(t)\frac{d}{dt}\left(A(t)^{-1}\right).
% \end{eqnarray}
% for a matrix $A(t)$.
% \end{rem}

\end{subsection}

\begin{subsection}{The Lax pair and conserved quantities of the delay Toda lattice equation under the periodic boundary condition}

Now, we introduce the $N$-periodic boundary condition $\psi_{n+N}(t)=\psi_{n}(t)$ in the space variable $n$.
By defining 
\begin{eqnarray}
    \hat{\psi}(t)=(\psi_{0}(t),\psi_{1}(t),\psi_{2}(t),\cdots,\psi_{N-1}(t))^{T},
\end{eqnarray}
equations (\ref{dlyTLLaxM}) and (\ref{dlyTLLaxL2}) in the case of $n=0,1,\cdots,N-1$ can be expressed in the following form:
\begin{eqnarray}
    -\frac{d}{dt}\hat{\psi}(t)+\mathcal{M}(t)\hat{\psi}(t)=\lambda_{1}\hat{\psi}(t),\label{dlyTLLaxMatrix1}\\
    \mathcal{L}_{2}(t)\hat{\psi}(t-2\tau)=\lambda_{2}\hat{\psi}(t),\label{dlyTLLaxMatrix2}
\end{eqnarray}
where $\mathcal{M}(t)$ and $\mathcal{L}_{2}(t)$ are described by matrices.
For example, they are defined as follows in the case of $\beta=0$:
\begin{eqnarray}
\fl \mathcal{M}(t)=
\left(
\begin{array}{cccccc}
T_{0}(t) & 1         & 0         &0         & \cdots & 0 \\
0         & T_{1}(t) & 1         &0         & \cdots & 0 \\
0         & 0         & T_{2}(t) &1         & \cdots & 0 \\
\vdots    & \vdots    & \ddots    &\ddots    & \ddots & \vdots \\
0         & 0         & \cdots    &0         & T_{N-2}(t) &1 \\
1         & 0         & \cdots    &0         & 0 & T_{N-1}(t) \\
\end{array}
\right),\label{dlyTLmatrixM}\\
\fl \mathcal{L}_{2}(t)=
\left(
\begin{array}{ccccc}
-1 & 0 & 0 & \cdots & 2\tau(1+V_{0}(t-2\tau)) \\
2\tau(1+V_{1}(t-2\tau)) & -1 & 0 & \cdots & 0 \\
0 & 2\tau(1+V_{2}(t-2\tau)) & -1 & \cdots & 0 \\
\vdots & \vdots & \ddots & \ddots & \vdots \\
0 & 0 & \cdots & 2\tau(1+V_{N-1}(t-2\tau)) & -1 \\
\end{array}
\right).\nonumber\\\label{dlyTLmatrixL2}
\end{eqnarray}

Now, by using the above Lax pair, we construct conserved quantities as functions of $t$ under the $N$-periodic boundary condition in the spatial variable $n$.
The compatibility condition of equations (\ref{dlyTLLaxMatrix1}) and (\ref{dlyTLLaxMatrix2}) is
\begin{eqnarray}
    \frac{d}{dt}\mathcal{L}_{2}(t)=\mathcal{M}(t)\mathcal{L}_{2}(t)-\mathcal{L}_{2}(t)\mathcal{M}(t-2\tau).\label{dlyTLMatrix_compa1}
\end{eqnarray}
In the case of $\tau=0$, the function $\mathrm{Tr}\left(\mathcal{L}_{2}(m)^{k}\right)$ are conserved quantities.
On the other hand, when $\tau\neq0$, equation (\ref{dlyTLMatrix_compa1}) cannot be used directly to construct conserved quantities, because the delay $\tau$ is contained and the right side of the equation cannot be expressed in terms of commutators.
It is necessary to apply the delay-differential limit to equation (\ref{dlydisTLMatrix_compa2}).

To consider the delay-differential limit of (\ref{dlydisTLMatrix_compa2}), we impose $\alpha\geq0$ and define a new discrete linear operator 
\begin{eqnarray}
    \bar{\mathcal{N}}^{*}(m)&=&\delta^{\alpha+1}\bar{\mathcal{N}}(m)\nonumber\\
    &=&\delta^{\alpha+1}\bar{\mathcal{L}}_{1}(m)\bar{\mathcal{L}}_{1}(m-1)\cdots \bar{\mathcal{L}}_{1}(m-\alpha)\bar{\mathcal{L}}_{2}(m)^{-1}.\label{dlydisTLop1}
\end{eqnarray}
% where $\bar{\mathcal{L}}_{1}(m)$ and $\bar{\mathcal{L}}_{2}(m)$ are defined by (\ref{dlydisTLLaxMatrixL1}) and (\ref{dlydisTLLaxMatrixL2}).
% Equation (\ref{dlydisTLop1}) can be rewritten by
This can be rewritten by
\begin{eqnarray}
    \fl \bar{\mathcal{N}}^{*}(m)=(I+\delta \bar{\mathcal{M}}(m))(I+\delta \bar{\mathcal{M}}(m-1))\cdots(I+\delta \bar{\mathcal{M}}(m-\alpha))\bar{\mathcal{L}}_{2}(m)^{-1},\label{dlydisTLop2}
\end{eqnarray}
where $\bar{\mathcal{M}}(m)$ is defined by
\begin{eqnarray}
    \bar{\mathcal{L}}_{1}(m)=\frac{1}{\delta}I+\bar{\mathcal{M}}(m)
\end{eqnarray} 
and $I$ is the identity matrix.
Since $\mu=1/\delta$, we can show that $\bar{\mathcal{M}}(m)$ and $\bar{\mathcal{L}}_{2}(m)$ are reduced to $\mathcal{M}(t)$ and $\mathcal{L}_{2}(t)$ as $\delta\to0$ respectively.
Applying the delay-differential limit (\ref{dlydiffLimitTL}) to (\ref{dlydisTLop2}), we obtain the following $\mathcal{N}^{*}(t)$ as the limit of $\bar{\mathcal{N}}^{*}(m)$:
\begin{eqnarray}
    \mathcal{N}^{*}(t)=E(t)\mathcal{L}_{2}(t)^{-1},\label{dlyTLop2}
\end{eqnarray}
where
\begin{eqnarray}
    \fl E(t)=&I+\int_{0}^{2\tau}dx_{1} \mathcal{M}(t-x_{1})+\int_{0}^{2\tau}dx_{2}\int_{0}^{x_{2}}dx_{1} \mathcal{M}(t-x_{1})\mathcal{M}(t-x_{2})\nonumber\\
    \fl &+\int_{0}^{2\tau}dx_{3}\int_{0}^{x_{3}}dx_{2}\int_{0}^{x_{2}}dx_{1} \mathcal{M}(t-x_{1})\mathcal{M}(t-x_{2})\mathcal{M}(t-x_{3})+\cdots.\label{dlyTL_E_2}
\end{eqnarray}
% Using (\ref{dlyTLcompa}) and
% \begin{eqnarray}
%    \frac{d}{dt}E(t)=E(t)M(t+2\tau)-\mathcal{M}(t)E(t)\,,
% \end{eqnarray}
These calculations can be carried out in the same way as in \ref{sec_derivation}.
On the other hand, as the delay-differential limit (\ref{dlydiffLimitTL}) of equation (\ref{dlydisTLMatrix_compa2}), we obtain
\begin{eqnarray}
    \frac{d}{dt}\mathcal{N}^{*}(t)=\mathcal{M}(t)\mathcal{N}^{*}(t)-\mathcal{N}^{*}(t)\mathcal{M}(t).\label{dlyTLMatrix_compa2}
\end{eqnarray}
Therefore, defining $H_{k}(t)=\mathrm{Tr}\left(\mathcal{N}^{*}(t)^{k}\right)$, we obtain
\begin{eqnarray}
    \frac{d}{dt}H_{k}(t)
    =\mathrm{Tr}\left(k\mathcal{N}^{*}(t)^{k-1}\frac{d}{dt}\mathcal{N}^{*}(t)\right)
    =0.
\end{eqnarray}
Thus $H_{k}(t)\ (k=1,2,\cdots)$ are conserved quantities of the delay Toda lattice equation.

We remark that $E(t)$, which is a part of the conserved quantities $H_{k}(t)$, is rewritten by 
\begin{eqnarray}
    E(t)=T'\left(\exp\left(\int_{0}^{2\tau}\mathcal{M}(t-s)ds\right)\right).\label{dlyTL_E}
\end{eqnarray}
$T'$ is the time-ordered product defined by 
\begin{eqnarray}
    T'\left(A(t_{1})A(t_{2})\cdots A(t_{i})\right)=A(t_{\sigma(1)})A(t_{\sigma(2)})\cdots A(t_{\sigma(i)}),
\end{eqnarray}
where $\sigma$ is a permutation that satisfies $t_{\sigma(1)}\geq t_{\sigma(2)}\geq\cdots\geq t_{\sigma(i)}$.
$T'$ sorts $A(t_1),\cdots,A(t_i)$ in decreasing order of time.
This calculation can be carried out in the same way as in \ref{sec_derivation}.

% The fact that $\bar{\mathcal{L}}_{1}$ is acted on $\alpha$ times in (\ref{dlydisTLop1}) and the delay-differential limit $\left|\alpha\right|\to\infty$ (\ref{dlydiffLimitTL}) cause the exponential structure (\ref{dlyTL_E}) in the conserved quantities $H_{k}(t)$.
% This is an effect caused by the delay.

\end{subsection}

\end{section}

\begin{section}{Relationships between delay soliton equations and delay Painlev\'e equations}
\label{sec_reduction}

% We move on to a discussion of relationships between delay soliton and delay Painlev\'e equations.
In this section, we show the relationships between delay soliton equations and delay Painlev\'e equations.
More precisely, it is shown that reductions of delay soliton equations lead to the already known delay Painlev\'e equations.
In addition, we consider discrete analogues of the delay Painlev\'e I\hspace{-1pt}I and I\hspace{-1pt}I\hspace{-1pt}I equations by reductions of discrete-time delay soliton equations.

\begin{subsection}{A reduction of the delay Lotka-Volterra equation}

We show that the spatial $2$-periodic reduction of the delay LV equation leads to the delay Painlev\'e I\hspace{-1pt}I equation~\cite{Gram}.

When $\beta=0$, transforming $u_{n}(t)\to\gamma_{n}e^{\omega t}u_{n}(t)\ (\omega=\mathrm{const}.)$ to the delay LV equation (\ref{dlyLVn}) yields
\begin{eqnarray}
    \label{dlylv_nl}
    \fl\frac{d}{dt}\log\frac{u_{n}(t+\tau)}{u_{n-1}(t-\tau)}
    =\gamma_{n+1}&e^{\omega(t+\tau)}u_{n+1}(t+\tau)-\gamma_{n}e^{\omega(t+\tau)}u_{n}(t+\tau)\nonumber\\
    \fl&-\gamma_{n-1}e^{\omega(t-\tau)}u_{n-1}(t-\tau)+\gamma_{n-2}e^{\omega(t-\tau)}u_{n-2}(t-\tau)\,.
\end{eqnarray}
The bilinear form of equation (\ref{dlylv_nl}) is given by
\begin{eqnarray}
    \label{dlylv}
    \fl D_{t}f_{n}(t+\tau)\cdot f_{n-1}(t-\tau)
    -\gamma_{n}e^{\omega t}f_{n+1}(t+\tau)f_{n-2}(t-\tau)\nonumber\\
    +A(t)f_{n}(t+\tau)f_{n-1}(t-\tau)=0
\end{eqnarray}
via the transformation
\begin{equation}
    u_{n}(t)=\frac{f_{n+1}(t+\tau)f_{n-2}(t-\tau)}{f_{n}(t+\tau)f_{n-1}(t-\tau)}
\end{equation}
and with an arbitrary function $A(t)$.

Now, in order to derive the delay Painlev\'e I\hspace{-1pt}I equation, we impose the $2$-periodic reduction in the spatial variable $n$ on $f_{n}(t)$.
Setting $F(t)=f_{0}(t)$, $G(t)=f_{1}(t)$, we obtain the following equations from (\ref{dlylv}):
\begin{eqnarray}
    \label{dlyp2_3}
    \fl D_{t}F(t+\tau)\cdot G(t-\tau)
    -\gamma_0e^{\omega t}F(t-\tau)G(t+\tau)
    +A(t)F(t+\tau)G(t-\tau)
    =0\,,\\
    \label{dlyp2_4}
    \fl D_{t}F(t-\tau)\cdot G(t+\tau)
    +\gamma_1e^{\omega t}F(t+\tau)G(t-\tau)
    -A(t)F(t-\tau)G(t+\tau)
    =0\,.
\end{eqnarray}
These equations (\ref{dlyp2_3}) and (\ref{dlyp2_4}) are the bilinear form of the delay Painlev\'e I\hspace{-1pt}I equation~\cite{Carstea}.
Through the transformation $v=G/F$, equations (\ref{dlyp2_3}) and (\ref{dlyp2_4}) lead to the nonlinear form of the delay Painlev\'e I\hspace{-1pt}I equation~\cite{Gram}:
\begin{equation}
    \label{dlyp2_nl}
    \frac{d}{dt}(\underline{v}\overline{v})=e^{\omega t}(\gamma_1\underline{v}^2-\gamma_0\overline{v}^2)\,,\qquad
    \overline{v}=v(t+\tau)\,,\qquad
    \underline{v}=v(t-\tau)\,.
\end{equation}
Note that~\cite{Gram,Carstea} show the delay Painlev\'e I\hspace{-1pt}I equation (\ref{dlyp2_nl}) and its bilinear form (\ref{dlyp2_3}) and (\ref{dlyp2_4}) are reduced to the Painlev\'e I\hspace{-1pt}I equation
\begin{equation}
    \label{p2_nl}
    \frac{d^2}{dt^2}w=2w^3+tw-a
\end{equation}
and its bilinear form
\begin{eqnarray}
    \label{p2_3}
    D_{t}^{3}F(t)\cdot G(t)-tD_{t}F(t)\cdot G(t)-aF(t)G(t)=0\,,\\
    \label{p2_4}
    D_{t}^{2}F(t)\cdot G(t)=0
\end{eqnarray}
by the continuum limit
\begin{equation*}
    \fl\gamma_0=-\frac{1}{2\tau}+\frac{\tau^2a}{3}\,,\quad
    \gamma_1=-\frac{1}{2\tau}-\frac{\tau^2a}{3}\,,\quad
    \omega=\frac{\tau^2}{3}\,,\quad
    A(t)=-\frac{e^{\omega t}}{2\tau}\,,\quad
    w=(\log v)'\,,\quad
    \tau\to0\,.
\end{equation*}

Although we used bilinear forms in the above derivation, we do not need them.
Let us directly derive the nonlinear form of the delay Painlev\'e I\hspace{-1pt}I equation (\ref{dlyp2_nl}) from the nonlinear delay LV equation (\ref{dlylv_nl}).
We first apply the variable transformation
\begin{equation}
    u_{n}(t)=\frac{\nu_{n}(t+\tau)}{\nu_{n-2}(t-\tau)}\,,
\end{equation}
and then obtain the following equation from (\ref{dlylv_nl}):
\begin{equation}
    \label{dlylv_nl2}
    \frac{d}{dt}\log\frac{\nu_{n}(t+\tau)}{\nu_{n-1}(t-\tau)}
    =\gamma_{n+1}e^{\omega t}\frac{\nu_{n+1}(t+\tau)}{\nu_{n-1}(t-\tau)}-\gamma_{n}e^{\omega t}\frac{\nu_{n}(t+\tau)}{\nu_{n-2}(t-\tau)}\,.
\end{equation}
Here, we impose the following reduction condition using the new dependent variable $v(t)$:
\begin{equation}
    \nu_{n+2}(t)=\nu_{n}(t)\,,\qquad
    \nu_{0}(t)=v(t)\,,\qquad
    \nu_{1}(t)=\frac{1}{v(t)}\,.
\end{equation}
By this reduction, equation (\ref{dlylv_nl2}) leads to the delay Painlev\'e I\hspace{-1pt}I equation (\ref{dlyp2_nl}).

\end{subsection}

\begin{subsection}{A reduction of the discrete-time delay Lotka-Volterra equation}

We consider the spatial $2$-periodic reduction of the discrete-time delay LV equation.

When $\beta=0$, by applying to the transformation $u_{n}^{m}\to\gamma_{n}q^{m}u_{n}^{m}\ (q=\mathrm{const}.)$ to the discrete-time delay LV equation (\ref{dlydisLVn}), we obtain
\begin{equation}
    \label{dlydislv_nl}
    \frac{u_{n}^{m+1+\alpha}u_{n-1}^{m}}{u_{n}^{m+\alpha}u_{n-1}^{m+1}}
    =\frac{(1+\delta\gamma_{n+1}q^{m+\alpha}u_{n+1}^{m+\alpha})(1+\delta\gamma_{n-2}q^{m+1}u_{n-2}^{m+1})}
    {(1+\delta\gamma_{n}q^{m+\alpha}u_{n}^{m+\alpha})(1+\delta\gamma_{n-1}q^{m+1}u_{n-1}^{m+1})}\,.
\end{equation}
The bilinear form of equation (\ref{dlydislv_nl}) is given by
\begin{eqnarray}
    \label{dlydislv}
    B_{m}f_{n}^{m+1+\alpha}f_{n-1}^{m}
    -\delta\gamma_{n}q^{m}f_{n+1}^{m+\alpha}f_{n-2}^{m+1}
    -f_{n}^{m+\alpha}f_{n-1}^{m+1}=0
\end{eqnarray}
via the transformation
\begin{equation}
    u_{n}^{m}=\frac{f_{n+1}^{m+\alpha}f_{n-2}^{m+1}}{f_{n}^{m+\alpha}f_{n-1}^{m+1}}
\end{equation}
and with an arbitrary function $B_{m}$.

Now, we impose the $2$-periodic reduction $f_{n+2}^{m}=f_{n}^{m}$.
Setting $F_{m}=f_{0}^{m}$, $G_{m}=f_{1}^{m}$, we obtain the following equations from (\ref{dlydislv}):
\begin{eqnarray}
    \label{dlydisp2_1}
    B_{m}F_{m+1+\alpha}G_{m}
    -\delta\gamma_{0}q^{m}F_{m+1}G_{m+\alpha}
    -F_{m+\alpha}G_{m+1}
    =0\,,\\
    \label{dlydisp2_2}
    B_{m}F_{m}G_{m+1+\alpha}
    -\delta\gamma_{1}q^{m}F_{m+\alpha}G_{m+1}
    -F_{m+1}G_{m+\alpha}
    =0\,.
\end{eqnarray}
Through the transformation $v_{m}=G_{m}/F_{m}$, equations (\ref{dlydisp2_1}) and (\ref{dlydisp2_2}) become
\begin{equation}
    \label{dlydisp2_nl}
    \frac{\overline{v_{m+1}}}{v_{m}}
    =\frac{\overline{v_{m}}+\delta\gamma_{1}q^{m}v_{m+1}}{v_{m+1}+\delta\gamma_{0}q^{m}\overline{v_{m}}}\,,\qquad
    \overline{v_{m}}=v_{m+\alpha}\,.
\end{equation}
We can check this equation exhibits singularity confinement behaviors.
It leads to the already known discrete analogues of the Painlev\'e I\hspace{-1pt}I equation when $\alpha=2,3$, and leads to the delay Painlev\'e I\hspace{-1pt}I equation (\ref{dlyp2_nl}) in the small limit of $\delta$ as discussed in the next subsection.
Thus, equation (\ref{dlydisp2_nl}) can be considered as a discrete analogue of the delay Painlev\'e I\hspace{-1pt}I equation.
On the other hands, equation (\ref{dlydisp2_nl}) is also considered as a higher order analogue of the discrete Painlev\'e I\hspace{-1pt}I equation.

We can also directly derive the equation (\ref{dlydisp2_nl}) from the nonlinear form of the discrete-time delay LV equation (\ref{dlydislv_nl}).
Applying the variable transformation
\begin{equation}
    u_{n}^{m}=\frac{\nu_{n}^{m+\alpha}}{\nu_{n-2}^{m+1}}
\end{equation}
and the reduction condition
\begin{equation}
    \nu_{n+2}^{m}=\nu_{n}^{m}\,,\qquad
    \nu_{0}^{m}=v_{m}\,,\qquad
    \nu_{1}^{m}=\frac{1}{v_{m}}
\end{equation}
to equation (\ref{dlydislv_nl}), we obtain equation (\ref{dlydisp2_nl}).

\end{subsection}

\begin{subsection}{Special cases of equation (\ref{dlydisp2_nl})}
\label{subsec_dlydisp2_nl}

Equation (\ref{dlydisp2_nl}) leads to the already known discrete Painlev\'e I\hspace{-1pt}I equations for specific choices of $\alpha$.
For example, setting $\alpha=2,\ \gamma_0=\gamma_1=1$, we obtain the following equation from (\ref{dlydisp2_nl}):
\begin{equation}
    \frac{v_{m+3}}{v_{m}}
    =\frac{v_{m+2}+\delta q^{m}v_{m+1}}{v_{m+1}+\delta q^{m}v_{m+2}}\,.
\end{equation}
Putting $x_{m}=v_{m+2}/v_{m+1}$, we obtain
\begin{equation}
    \label{disp2_nl}
    x_{m+1}x_{m-1}
    =\frac{1}{x_{m}}\frac{x_{m}+\delta q^{m}}{1+\delta q^{m}x_{m}}\,.
\end{equation}
This equation (\ref{disp2_nl}) is the multiplicative discrete Painlev\'e I\hspace{-1pt}I equation introduced in~\cite{Ramani2}.
Next, setting $\alpha=3$ and $x_{m}=v_{m+3}/v_{m+1}$, we obtain the following equation from (\ref{dlydisp2_nl}):
\begin{equation}
    \label{disp2_2_nl}
    x_{m+1}x_{m-1}
    =\frac{x_{m}+\delta q^{m}}{1+\delta q^{m}x_{m}}\,.
\end{equation}
This equation (\ref{disp2_2_nl}) is also known as the discrete Painlev\'e I\hspace{-1pt}I equation~\cite{Ramani2}.
%(Note that similar derivations were carried out in~\cite{Gram2} by using reductions of the discrete \mkdv.)

On the other hand, equation (\ref{dlydisp2_nl}) leads to the delay Painlev\'e I\hspace{-1pt}I equation (\ref{dlyp2_nl}) in the delay-differential limit:
\begin{equation}
    \delta=e^{\omega\tau}\delta'\,,\quad
    m\delta'=t\,,\quad
    \alpha\delta'=2\tau\,,\quad
    q=1+\delta'\omega\,,\quad
    %B_{m}=1+\delta' A(t+\tau)\,,\quad
    \delta'\to0\,.
\end{equation}
%We can claim that equation (\ref{dlydisp2_nl}) extends some discrete and delay Painlev\'e I\hspace{-1pt}I equations.

\end{subsection}

\begin{subsection}{A reduction of the delay Toda lattice equation}

We show that the spatial $2$-periodic reduction of the delay Toda lattice equation leads to the delay Painlev\'e I\hspace{-1pt}I\hspace{-1pt}I equation~\cite{Gram}.

When $\beta=0$, transforming $1+V_{n}(t)\to\gamma_{n}e^{\omega t}(1+V_{n}(t))\ (\omega=\mathrm{const}.)$ to the delay Toda lattice equation (\ref{dlyTLn}) yields
\begin{eqnarray}
    \label{dlytl_nl}
    \fl \frac{d}{dt}\log \frac{1+V_{n}(t+\tau)}{1+V_{n}(t-\tau)}
    =2\tau\{&\gamma_{n+1}e^{\omega(t+\tau)}(1+V_{n+1}(t+\tau))+\gamma_{n-1}e^{\omega(t-\tau)}(1+V_{n-1}(t-\tau))\nonumber\\
    \fl&-\gamma_{n}e^{\omega(t+\tau)}(1+V_{n}(t+\tau))-\gamma_{n}e^{\omega(t-\tau)}(1+V_{n}(t-\tau))\}\,.
\end{eqnarray}
The bilinear form of equation (\ref{dlytl_nl}) is given by
\begin{eqnarray}
    \label{dlytl}
    \fl D_{t}f_{n}(t+\tau)\cdot f_{n}(t-\tau)
    -2\tau\gamma_{n}e^{\omega t}f_{n+1}(t+\tau)f_{n-1}(t-\tau)\nonumber\\
    +A(t)f_{n}(t+\tau)f_{n}(t-\tau)=0
\end{eqnarray}
via the transformation
\begin{equation}
    1+V_{n}(t)=\frac{f_{n+1}(t+\tau)f_{n-1}(t-\tau)}{f_{n}(t+\tau)f_{n}(t-\tau)}
\end{equation}
and with an arbitrary function $A(t)$.

Now, in order to derive the delay Painlev\'e I\hspace{-1pt}I\hspace{-1pt}I equation, we impose the $2$-periodic reduction in the spatial variable $n$ on $f_{n}(t)$.
Setting $F(t)=f_{0}(t)$, $G(t)=f_{1}(t)$, we obtain the following equations from (\ref{dlytl}):
\begin{eqnarray}
    \label{dlyp3_3}
    \fl D_{t}F(t+\tau)\cdot F(t-\tau)
    -2\tau\gamma_0e^{\omega t}G(t+\tau)G(t-\tau)
    +A(t)F(t+\tau)F(t-\tau)
    =0\,,\\
    \label{dlyp3_4}
    \fl D_{t}G(t+\tau)\cdot G(t-\tau)
    -2\tau\gamma_1e^{\omega t}F(t+\tau)F(t-\tau)
    +A(t)G(t+\tau)G(t-\tau)
    =0\,.
\end{eqnarray}
These equations (\ref{dlyp3_3}) and (\ref{dlyp3_4}) are the bilinear form of the delay Painlev\'e I\hspace{-1pt}I\hspace{-1pt}I equation~\cite{Carstea}.
Through the transformation $v=G/F$, equations (\ref{dlyp3_3}) and (\ref{dlyp3_4}) lead to the nonlinear form of the delay Painlev\'e I\hspace{-1pt}I\hspace{-1pt}I equation~\cite{Gram}:
\begin{equation}
    \label{dlyp3_nl}
    \frac{d}{dt}\log\frac{\overline{v}}{\underline{v}}=2\tau e^{\omega t}\left(\frac{\gamma_1}{\underline{v}\overline{v}}-\gamma_0\underline{v}\overline{v}\right)\,,\qquad
    \overline{v}=v(t+\tau)\,,\qquad
    \underline{v}=v(t-\tau)\,.
\end{equation}
Note that~\cite{Gram,Carstea} show the delay Painlev\'e I\hspace{-1pt}I\hspace{-1pt}I equation (\ref{dlyp3_nl}) and its bilinear form (\ref{dlyp3_3}) and (\ref{dlyp3_4}) are reduced to a particular form of the Painlev\'e I\hspace{-1pt}I\hspace{-1pt}I equation
\begin{equation}
    \label{p3_nl}
    \frac{d^2v}{dt^2}=\frac{1}{v}\left(\frac{dv}{dt}\right)^2+e^{\omega t}\left(-\gamma_0v^3+\frac{\gamma_1}{v}\right)
\end{equation}
and its bilinear form
\begin{eqnarray}
    \label{p3_3}
    D_{t}^{2}F(t)\cdot F(t)
    -\gamma_0e^{\omega t}G(t)G(t)
    +A'(t)F(t)F(t)
    =0\,,\\
    \label{p3_4}
    D_{t}^{2}G(t)\cdot G(t)
    -\gamma_1e^{\omega t}F(t)F(t)
    +A'(t)G(t)G(t)
    =0
\end{eqnarray}
by the continuum limit $A(t)=2\tau A'(t),\ \tau\to0$.
The general form of the Painlev\'e I\hspace{-1pt}I\hspace{-1pt}I equation is
\begin{equation}
    \label{p3_general_nl}
    \frac{d^2v}{dt^2}
    =\frac{1}{v}\left(\frac{dv}{dt}\right)^2
    +e^{2t}\left(\kappa_0v^3+\frac{\kappa_1}{v}\right)
    +e^{t}\left(\kappa_2v^2+\kappa_3\right)\,.\label{PIIIexp}
\end{equation}
Equation (\ref{p3_nl}) corresponds to the case $\kappa_2=\kappa_3=0$. 
By applying the independent variable transformation $t=\log s$ to equation (\ref{PIIIexp}), we obtain the following usual form of the Painlev\'{e} III equation:
\begin{eqnarray}
        \frac{d^{2}v}{ds^{2}}=\frac{1}{v}\left(\frac{dv}{ds}\right)^{2}-\frac{1}{s}\frac{dv}{ds}+\frac{1}{s}\left(\kappa_{2}v^{2}+\kappa_{3}\right)+\kappa_{0}v^{3}+\frac{\kappa_{1}}{v}.
\end{eqnarray}    

Although we used bilinear forms in the above derivation, we do not need them.
Let us directly derive the nonlinear form of the delay Painlev\'e I\hspace{-1pt}I\hspace{-1pt}I equation (\ref{dlyp3_nl}) from the nonlinear delay Toda lattice equation (\ref{dlytl_nl}).
We first apply the variable transformation
\begin{equation}
    1+V_{n}(t)=\frac{\nu_{n}(t+\tau)}{\nu_{n-1}(t-\tau)}\,,
\end{equation}
and then obtain the following equation from (\ref{dlytl_nl}):
\begin{equation}
    \label{dlytl_nl2}
    \frac{d}{dt}\log\frac{\nu_{n}(t+\tau)}{\nu_{n}(t-\tau)}
    =2\tau\gamma_{n+1}e^{\omega t}\frac{\nu_{n+1}(t+\tau)}{\nu_{n}(t-\tau)}-2\tau\gamma_{n}e^{\omega t}\frac{\nu_{n}(t+\tau)}{\nu_{n-1}(t-\tau)}\,.
\end{equation}
Here, we impose the following reduction condition using the new dependent variable $v(t)$:
\begin{equation}
    \nu_{n+2}(t)=\nu_{n}(t)\,,\qquad
    \nu_{0}(t)=v(t)\,,\qquad
    \nu_{1}(t)=\frac{1}{v(t)}\,.
\end{equation}
By this reduction, equation (\ref{dlytl_nl2}) leads to the delay Painlev\'e I\hspace{-1pt}I\hspace{-1pt}I equation (\ref{dlyp3_nl}).

\end{subsection}

\begin{subsection}{A reduction of the discrete-time delay Toda lattice equation}

We consider the spatial $2$-periodic reduction of the discrete-time delay Toda lattice equation.

When $\beta=0$, by applying to the transformation $1+V_{n}^{m}\to\gamma_{n}q^{m}(1+V_{n}^{m})\ (q=\mathrm{const}.),\ c\delta/(1-c\delta)\to c\delta$ to the discrete-time delay Toda lattice equation (\ref{dlydisTLn}), we obtain
\begin{equation}
    \label{dlydistl_nl}
    \fl \frac{(1+V_{n}^{m+1+\alpha})(1+V_{n}^{m-1})}{(1+V_{n}^{m+\alpha})(1+V_{n}^{m})}=\frac{(1+c\delta\gamma_{n+1}q^{m+\alpha}(1+V_{n+1}^{m+\alpha}))(1+c\delta\gamma_{n-1}q^{m}(1+V_{n-1}^{m}))}{(1+c\delta\gamma_{n}q^{m+\alpha}(1+V_{n}^{m+\alpha}))(1+c\delta\gamma_{n}q^{m}(1+V_{n}^{m}))}\,.
\end{equation}
The bilinear form of equation (\ref{dlydistl_nl}) is given by
\begin{eqnarray}
    \label{dlydistl}
    B_{m}f_{n}^{m+1+\alpha}f_{n}^{m-1}
    -c\delta\gamma_{n}q^{m}f_{n+1}^{m+\alpha}f_{n-1}^{m}
    -f_{n}^{m+\alpha}f_{n}^{m}=0\,.
\end{eqnarray}
via the transformation
\begin{equation}
    1+V_{n}^{m}=\frac{f_{n+1}^{m+\alpha}f_{n-1}^{m}}{f_{n}^{m+\alpha}f_{n}^{m}}
\end{equation}
and with an arbitrary function $B_{m}$.

Now, we impose the $2$-periodic reduction $f_{n+2}^{m}=f_{n}^{m}$.
Setting $F_{m}=f_{0}^{m}$, $G_{m}=f_{1}^{m}$, we obtain the following equations from (\ref{dlydistl}):
\begin{eqnarray}
    \label{dlydisp3_1}
    B_{m}F_{m+1+\alpha}F_{m-1}
    -c\delta\gamma_{0}q^{m}G_{m+\alpha}G_{m}
    -F_{m+\alpha}F_{m}
    =0\,,\\
    \label{dlydisp3_2}
    B_{m}G_{m+1+\alpha}G_{m-1}
    -c\delta\gamma_{1}q^{m}F_{m+\alpha}F_{m}
    -G_{m+\alpha}G_{m}
    =0\,.
\end{eqnarray}
Through the transformation $v_{m}=G_{m}/F_{m}$, equations (\ref{dlydisp3_1}) and (\ref{dlydisp3_2}) become
\begin{equation}
    \label{dlydisp3_nl}
    \overline{v_{m+1}}v_{m-1}
    =\frac{\overline{v_{m}}v_{m}+c\delta\gamma_1q^{m}}{1+c\delta\gamma_{0}q^{m}\overline{v_{m}}v_{m}}\,,\qquad
    \overline{v_{m}}=v_{m+\alpha}\,.
\end{equation}
We can check this equation exhibits singularity confinement behaviors.
It leads to the already known discrete analogues of the Painlev\'e I\hspace{-1pt}I\hspace{-1pt}I equation when $\alpha=0$, and leads to the delay Painlev\'e I\hspace{-1pt}I\hspace{-1pt}I equation (\ref{dlyp3_nl}) in the small limit of $\delta$ as discussed in the next subsection.
Thus, equation (\ref{dlydisp3_nl}) can be considered as a discrete analogue of the delay Painlev\'e I\hspace{-1pt}I\hspace{-1pt}I equation.
On the other hands, equation (\ref{dlydisp3_nl}) is also considered as a higher order analogue of the discrete Painlev\'e I\hspace{-1pt}I\hspace{-1pt}I equation.

We can also directly derive equation (\ref{dlydisp3_nl}) from the nonlinear form of the discrete-time delay Toda lattice equation (\ref{dlydistl_nl}).
Applying the variable transformation
\begin{equation}
    1+V_{n}^{m}=\frac{\nu_{n}^{m+\alpha}}{\nu_{n-1}^{m}}
\end{equation}
and the reduction condition 
\begin{equation}
    \nu_{n+2}^{m}=\nu_{n}^{m}\,,\qquad
    \nu_{0}^{m}=v_{m}\,,\qquad
    \nu_{1}^{m}=\frac{1}{v_{m}}
\end{equation}
to equation (\ref{dlydistl_nl}), we obtain equation (\ref{dlydisp3_nl}).

\end{subsection}

\begin{subsection}{Special cases of equation (\ref{dlydisp3_nl})}

Equation (\ref{dlydisp3_nl}) leads to the already known discrete Painlev\'e I\hspace{-1pt}I\hspace{-1pt}I equation for a specific choice of $\alpha$.
Setting $\alpha=0,\ \gamma_0=\gamma_1=1$, we obtain the following equation from (\ref{dlydisp3_nl}):
\begin{equation}
    \label{disp3_nl}
    v_{m+1}v_{m-1}
    =\frac{v_{m}v_{m}+c\delta q^{m}}{1+c\delta q^{m}v_{m}v_{m}}\,.
\end{equation}
This equation (\ref{disp3_nl}) is a particular form of the discrete Painlev\'e I\hspace{-1pt}I\hspace{-1pt}I equation~\cite{Ramani3}.

On the other hand, equation (\ref{dlydisp3_nl}) leads to the delay Painlev\'e I\hspace{-1pt}I\hspace{-1pt}I equation (\ref{dlyp3_nl}) in the delay-differential limit:
\begin{equation}
    \delta=e^{\omega\tau}\delta'\,,\quad
    m\delta'=t\,,\quad
    c=\alpha\delta'=2\tau\,,\quad
    q=1+\delta'\omega\,,\quad
    %B_{m}=1+\delta' A(t+\tau)\,,\quad
    \delta'\to0\,.
\end{equation}
%We can claim that equation (\ref{dlydisp3_nl}) extends some discrete and delay Painlev\'e I\hspace{-1pt}I\hspace{-1pt}I equations.

\end{subsection}

\end{section}

\begin{section}{Constructions of determinant solutions of delay Painlev\'e equations and their discrete analogues}
\label{sec_detsol}

As applications of reductions discussed in section \ref{sec_reduction}, we construct the $N$-soliton-type solutions of the autonomous versions of the delay Painlev\'e I\hspace{-1pt}I and I\hspace{-1pt}I\hspace{-1pt}I equations.
These solutions have determinant structures of size $N$.
These examples exist due to the inclusion of delay parameters in the equations.
Then, we also construct the Casorati determinant solution of a higher order analogue of the discrete Painlev\'e I\hspace{-1pt}I equation by direct calculation.

\begin{subsection}{The $N$-soliton-type solution of the autonomous version of the delay Painlev\'e I\hspace{-1pt}I equation}

By using the $2$-periodic reductions in section \ref{sec_reduction}, we can obtain the $N$-soliton-type determinant solutions of the autonomous case of the delay Painlev\'e I\hspace{-1pt}I equation (\ref{dlyp2_nl})
\begin{equation}
    \label{dlyp2_nl_auto}
    \frac{d}{dt}(\underline{v}\overline{v})=(\underline{v}^2-\overline{v}^2)\,,\qquad
    \overline{v}=v(t+\tau)\,,\qquad
    \underline{v}=v(t-\tau)\,.
\end{equation}
Equation (\ref{dlyp2_nl_auto}) is the case $\omega=0,\ \gamma_0=1,\ \gamma_1=1$ of the delay Painlev\'e I\hspace{-1pt}I equation (\ref{dlyp2_nl}).

As discussed in section \ref{sec_reduction}, the bilinear form of (\ref{dlyp2_nl_auto}) is
\begin{eqnarray}
    \label{dlyp2_auto_1}
    \fl D_{t}F(t+\tau)\cdot G(t-\tau)
    -F(t-\tau)G(t+\tau)
    +F(t+\tau)G(t-\tau)
    =0\,,\\
    \label{dlyp2_auto_2}
    \fl D_{t}F(t-\tau)\cdot G(t+\tau)
    +F(t+\tau)G(t-\tau)
    -F(t-\tau)G(t+\tau)
    =0\,,
\end{eqnarray}
where $v=G/F$.
These equations are obtained by the reduction
\begin{equation}
    \label{reduction_dlyLV_beta0}
    f_{n+2}(t)=f_{n}(t),\quad
    F(t)=f_{0}(t),\quad
    G(t)=f_{1}(t)
\end{equation}
of the delay LV equation (\ref{dlyLVb}) in the case of $\beta=0$:
\begin{eqnarray}
    \label{dlyLVb_beta0}
    \fl D_{t}f_{n}(t+\tau)\cdot f_{n-1}(t-\tau)-f_{n+1}(t+\tau)f_{n-2}(t-\tau)+f_{n}(t+\tau)f_{n-1}(t-\tau)=0\,.
\end{eqnarray}
The $N$-soliton solution of (\ref{dlyLVb_beta0}) is as follows~\cite{Nakata1}:
\begin{eqnarray}
    \label{dlyLVb_beta0_sol}
    f_{n}(t)
    =\det\left(\delta_{ij}+\frac{\Phi_j}{p_i-q_j}\right)_{1\leq i,j\leq N}\,,\\
    \Phi_i(n,t)
    =\epsilon_i\left(\frac{1+q_i}{1+p_i}\right)^{n} e^{(q_i-p_i)t}\,,\quad
    \frac{q_i}{p_i}
    =\left(\frac{1+q_i}{1+p_i}\right)^2 e^{2\tau(q_i-p_i)}\,,\nonumber
\end{eqnarray}
where $\epsilon_i,\ p_i,\ q_i$ are constatnts.
Applying the reduction (\ref{reduction_dlyLV_beta0}) to the solution (\ref{dlyLVb_beta0_sol}), i.e. imposing the constraint $p_i+q_i=-2$ on the solution (\ref{dlyLVb_beta0_sol}), we obtain
\begin{eqnarray}
    \label{dlyp2_auto_sol}
    F(t)=f_{0}(t)
    =\det\left(\delta_{ij}+\frac{\Phi_j}{2+p_i+p_j}\right)_{1\leq i,j\leq N}\,,\\
    G(t)=f_{1}(t)
    =\det\left(\delta_{ij}-\frac{\Phi_j}{2+p_i+p_j}\right)_{1\leq i,j\leq N}\,,\nonumber\\
    \Phi_i(n,t)
    =\epsilon_ie^{-2(1+p_i)t}\,,\quad
    1+p_i
    =-\tanh2\tau(1+p_i)\,.\nonumber
\end{eqnarray}
It satisfies the bilinear equations (\ref{dlyp2_auto_1}) and (\ref{dlyp2_auto_2}), and $v=G/F$ is the $N$-soliton-type solution of (\ref{dlyp2_nl_auto}).

\begin{rem}
    Putting $\eta_i=-2-2p_i$, the above solution (\ref{dlyp2_auto_sol}) becomes
    \begin{eqnarray}
        F(t)=f_{0}(t)
        =\det\left(\delta_{ij}-\frac{2\Phi_j}{\eta_i+\eta_j}\right)_{1\leq i,j\leq N}\,,\\
        G(t)=f_{1}(t)
        =\det\left(\delta_{ij}+\frac{2\Phi_j}{\eta_i+\eta_j}\right)_{1\leq i,j\leq N}\,,\nonumber\\
        \Phi_i(n,t)
        =\epsilon_ie^{\eta_it}\,,\quad
        -\frac{\eta_i}{2}
        =\tanh\tau\eta_i\,.\nonumber
    \end{eqnarray}
    When $h:=-2\tau$, the above solution for $N=1,2,3$ corresponds to the solution derived in~\cite{Berntson}.
\end{rem}
\begin{rem}
    \label{rem_matrix_size}
    Putting $\eta_i=\sqrt{-1}\xi_i/\tau$, the dispersion relation becomes
    \begin{equation}
        -\frac{\xi_i}{2\tau}=\tan\xi_i\,.
    \end{equation}
    This equation has an infinite number of solutions $\xi_i$ when $\tau\neq0$.
    Thus the matrix size $N$ can be arbitrarily large.
\end{rem}

\end{subsection}

\begin{subsection}{The $N$-soliton-type solution of the autonomous version of the delay Painlev\'e I\hspace{-1pt}I\hspace{-1pt}I equation}

Similarly to the above approach, we can construct the $N$-soliton-type determinant solutions of the autonomous case of the delay Painlev\'e I\hspace{-1pt}I\hspace{-1pt}I equation (\ref{dlyp3_nl})
\begin{equation}
    \label{dlyp3_nl_auto}
    \frac{d}{dt}\log\frac{\overline{v}}{\underline{v}}=2\tau\left(\frac{1}{\underline{v}\overline{v}}-\underline{v}\overline{v}\right)\,,\qquad
    \overline{v}=v(t+\tau)\,,\qquad
    \underline{v}=v(t-\tau)\,.
\end{equation}
Equation (\ref{dlyp3_nl_auto}) is the case $\omega=0,\ \gamma_0=1,\ \gamma_1=1$ of the delay Painlev\'e I\hspace{-1pt}I\hspace{-1pt}I equation (\ref{dlyp3_nl}).

As discussed in section \ref{sec_reduction}, the bilinear form of (\ref{dlyp3_nl_auto}) is
\begin{eqnarray}
    \label{dlyp3_auto_1}
    \fl D_{t}F(t+\tau)\cdot F(t-\tau)
    -2\tau G(t+\tau)G(t-\tau)
    +2\tau F(t+\tau)F(t-\tau)
    =0\,,\\
    \label{dlyp3_auto_2}
    \fl D_{t}G(t+\tau)\cdot G(t-\tau)
    -2\tau F(t+\tau)F(t-\tau)
    +2\tau G(t+\tau)G(t-\tau)
    =0\,,
\end{eqnarray}
where $v=G/F$.
These equations are obtained by the reduction
\begin{equation}
    \label{reduction_dlyTL_beta0}
    f_{n+2}(t)=f_{n}(t),\quad
    F(t)=f_{0}(t),\quad
    G(t)=f_{1}(t)
\end{equation}
of the delay Toda lattice equation (\ref{dlyTLb}) in the case of $\beta=0$:
\begin{eqnarray}
    \label{dlyTLb_beta0}
    \fl D_{t}f_{n}(t+\tau)\cdot f_{n}(t-\tau)
    -2\tau(f_{n+1}(t+\tau)f_{n-1}(t-\tau)-f_{n}(t+\tau)f_{n}(t-\tau))=0\,.
\end{eqnarray}
The $N$-soliton solution of (\ref{dlyTLb_beta0}) is as follows~\cite{Nakata1}:
\begin{eqnarray}
    \label{dlyTLb_beta0_sol}
    f_{n}(t)
    =\det\left(\delta_{ij}+\frac{\Phi_j}{p_i-q_j}\right)_{1\leq i,j\leq N}\,,\\
    \Phi_i(n,t)
    =\epsilon_i\left(\frac{q_i-2\tau}{p_i-2\tau}\right)^{n} e^{(p_i-q_i)t}\,,\quad
    \frac{q_i}{p_i}
    =\frac{q_i-2\tau}{p_i-2\tau}e^{2\tau(p_i-q_i)}\,,\nonumber
\end{eqnarray}
where $\epsilon_i,\ p_i,\ q_i$ are constatnts.
Applying the reduction (\ref{reduction_dlyTL_beta0}) to the solution (\ref{dlyTLb_beta0_sol}), i.e. imposing the constraint $p_i+q_i=4\tau$ on the solution (\ref{dlyTLb_beta0_sol}), we obtain
\begin{eqnarray}
    \label{dlyp3_auto_sol}
    F(t)=f_{0}(t)
    =\det\left(\delta_{ij}+\frac{\Phi_j}{-4\tau+p_i+p_j}\right)_{1\leq i,j\leq N}\,,\\
    G(t)=f_{1}(t)
    =\det\left(\delta_{ij}-\frac{\Phi_j}{-4\tau+p_i+p_j}\right)_{1\leq i,j\leq N}\,,\nonumber\\
    \Phi_i(n,t)
    =\epsilon_ie^{2(p_i-2\tau)t}\,,\quad
    \frac{2\tau}{2\tau-p_i}
    =\tanh2\tau(p_i-2\tau)\,.\nonumber
\end{eqnarray}
It satisfies the bilinear equations (\ref{dlyp3_auto_1}) and (\ref{dlyp3_auto_2}), and $v=G/F$ is the $N$-soliton-type solution of (\ref{dlyp3_nl_auto}).

\begin{rem}
    Putting $\eta_i=2p_i-4\tau$, the above solution (\ref{dlyp3_auto_sol}) becomes
    \begin{eqnarray}
        F(t)=f_{0}(t)
        =\det\left(\delta_{ij}+\frac{2\Phi_j}{\eta_i+\eta_j}\right)_{1\leq i,j\leq N}\,,\\
        G(t)=f_{1}(t)
        =\det\left(\delta_{ij}-\frac{2\Phi_j}{\eta_i+\eta_j}\right)_{1\leq i,j\leq N}\,,\nonumber\\
        \Phi_i(n,t)
        =\epsilon_ie^{\eta_it}\,,\quad
        1
        =-\frac{\eta_i}{4\tau}\tanh\tau\eta_i\,.\nonumber
    \end{eqnarray}
    The cases of $N=1,2,3$ of the above solution were derived in~\cite{Berntson}.
    When $z:=2\tau t$ and $h:=-4\tau^2$, the above solution for $N=1,2,3$ corresponds to the solution derived in~\cite{Berntson}.
    Similarly to Remark \ref{rem_matrix_size}, the matrix size $N$ can be arbitrarily large.
\end{rem}

\end{subsection}

\begin{subsection}{The Casorati determinant solution of a higher order analogue of the discrete Painlev\'e I\hspace{-1pt}I equation}

As for (non-autonomous) delay Painlev\'e equations, it is difficult to construct determinant solutions by the above reductions.
Thus we use direct calculation to construct the Casorati determinant solution of a higher order analogue of the discrete Painlev\'e I\hspace{-1pt}I equation.

It is known that the multiplicative discrete Painlev\'e I\hspace{-1pt}I equation~\cite{Ramani2}
\begin{equation}
    \label{disp2_nl_particular}
    x_{n+1}x_{n-1}
    =q^{-2(2N+1)}\frac{1}{x_{n}}\frac{x_{n}+\delta q^{n}}{1+\delta q^{n}x_{n}}
\end{equation}
has the Casorati determinant solution~\cite{Nakao}
\begin{eqnarray}
    \label{disp2_sol_1}
    x_{n}=-q^{-2N-n}\frac{\tau_{N+1}(n)\tau_{N}(n-1)}{\tau_{N+1}(n-1)\tau_{N}(n)}\,,\\
    \label{disp2_sol_2}
    \tau_N(n)=
    \left|
    \begin{array}{cccc}
        f_{n}     & f_{n+2}   & \cdots & f_{n+2N-2} \\
        f_{n-1}   & f_{n+1}   & \cdots & f_{n+2N-3} \\
        \vdots    & \vdots    & \ddots & \vdots     \\
        f_{n-N+1} & f_{n-N+3} & \cdots & f_{n+N-1}  \\
    \end{array}
    \right|\,,\\
    \label{disp2_sol_3}
    \delta f_{n+1}-f_{n}+\delta q^{-2n}f_{n-1}=0\,.
\end{eqnarray}
In other words, the following bilinear equations
\begin{eqnarray}
    \label{disp2_1_particular}
    \delta F_{n-2}G_{n+1}-q^{2N}F_{n-1}G_{n}+\delta q^{4N+2n}F_{n}G_{n-1}=0\,,\\
    \label{disp2_2_particular}
    \delta F_{n}G_{n+1}-q^{2N}F_{n+1}G_{n}+\delta q^{2n}F_{n+2}G_{n-1}=0\,,
\end{eqnarray}
have the Casorati determinant solution:
\begin{equation}
    F_{n}=\tau_{N}(n)\,,\qquad
    G_{n}=\tau_{N+1}(n)\,.
\end{equation}
By extending this fact, we obtain Theorem \ref{thm_detsol}.
% Inspired by this fact, we are interested in whether solutions of the delay-discrete Painlev\'e I\hspace{-1pt}I equation (\ref{dlydisp2_nl}) also have the Casorati determinant structure, because the case $\alpha=2$ of equation (\ref{dlydisp2_nl}) is the multiplicative discrete Painlev\'e I\hspace{-1pt}I (\ref{disp2_nl}) as shown in subsection \ref{subsec_dlydisp2_nl}.

% Although it is difficult to construct determinant solutions of (\ref{dlydisp2_nl}), in this paper, we construct the Casorati determinant solution of another type of delay analogue of the multiplicative discrete Painlev\'e I\hspace{-1pt}I equation.

\begin{thm}
\label{thm_detsol}
The following Casorati determinant
\begin{eqnarray}
    \label{dlydisp2_alt_FG}
    F_{n}=\tau_{N}(n)\,,\quad
    G_{n}=\tau_{N+1}(n)\,,\\
    \label{dlydisp2_alt_tau}
    \tau_N(n)=
    \left|
    \begin{array}{cccc}
        f_{n}     & f_{n+\alpha}     & \cdots & f_{n+(N-1)\alpha}     \\
        f_{n-1}   & f_{n+\alpha-1}   & \cdots & f_{n+(N-1)\alpha-1}   \\
        \vdots    & \vdots        & \ddots & \vdots             \\
        f_{n-N+1} & f_{n+\alpha-N+1} & \cdots & f_{n+(N-1)(\alpha-1)} \\
    \end{array}
    \right|\,,\\
    \label{dlydisp2_alt_disper}
    \delta f_{n+\alpha}-f_{n+\alpha-1}+\delta(-1)^{\alpha}q^{2(\alpha-1)(n+1)}f_{n}=0
\end{eqnarray}
gives a solution to the following bilinear equations:
\begin{eqnarray}
    \label{dlydisp2_alt_1_particular}
    \fl\delta F_{n-1}G_{n+\alpha}
    -q^{2N(\alpha-1)}F_{n}G_{n+\alpha-1}
    +\delta(-1)^{\alpha}q^{2N\alpha(\alpha-1)+2(\alpha-1)(n+1)}F_{n+\alpha-1}G_{n}
    =0\,,\\
    \label{dlydisp2_alt_2_particular}
    \fl\delta F_{n}G_{n+1}
    -q^{2N(\alpha-1)}F_{n+1}G_{n}
    +\delta(-1)^{\alpha}q^{2(\alpha-1)(n-\alpha+2)}F_{n+\alpha}G_{n-\alpha+1}
    =0\,.
\end{eqnarray}
\end{thm}

This theorem is proved by Pl\"ucker relations in \ref{sec_proof}.
In this theorem, the iteration step in the column direction in the determinant is generalised from 2 to $\alpha$.
Equations (\ref{dlydisp2_alt_1_particular}) and (\ref{dlydisp2_alt_2_particular}) are reduced to equations (\ref{disp2_1_particular}) and (\ref{disp2_2_particular}) when $\alpha=2$.
Thus we can consider they are the bilinear form of a higher order analogue of the multiplicative discrete Painlev\'e I\hspace{-1pt}I equation (\ref{disp2_nl}).
The nonlinear form of the bilinear equations (\ref{dlydisp2_alt_1_particular}) and (\ref{dlydisp2_alt_2_particular}) is
\begin{eqnarray}
    \label{dlydisp2_alt_nl_1_particular}
    \delta y_{n}=q^{2N(\alpha-1)}-a_n\frac{1}{x_{n+\alpha/2-1}\cdots x_{n-\alpha/2+1}}\,,\\
    \label{dlydisp2_alt_nl_2_particular}
    \delta x_{n}=q^{2N(\alpha-1)}-b_n x_{n}\left(y_{n+\alpha/2-1}\cdots y_{n-\alpha/2+1}\right)\frac{1}{x_{n+\alpha-1}\cdots x_{n-\alpha+1}}\,,\\
    a_n:=\delta(-1)^{\alpha}q^{2N\alpha(\alpha-1)+2(\alpha-1)(n-\alpha/2+1)}\,,\\
    b_n:=\delta(-1)^{\alpha}q^{2(\alpha-1)(n-\alpha+1)}\,.
\end{eqnarray}
where
\begin{equation}
    x_{n}=\frac{G_{n}F_{n-1}}{F_{n}G_{n-1}}\,,\quad
    y_{n}=\frac{G_{n+\alpha/2}F_{n-\alpha/2-1}}{F_{n-\alpha/2}G_{n+\alpha/2-1}}\,.
\end{equation}
Equations (\ref{dlydisp2_alt_nl_1_particular}) and (\ref{dlydisp2_alt_nl_2_particular}) are reduced to equation (\ref{disp2_nl_particular}) when $\alpha=2$.

It is not certain what differential equation is obtained by the continuum limit of this equation.
Additionally, it is unknown whether the higher order analogue of the discrete Painlev\'e I\hspace{-1pt}I equation (\ref{dlydisp2_nl}) also has determinant solutions similarly.
These issues remain subjects for future studies.

\end{subsection}

\end{section}

\begin{section}{Conclusions}
\label{sec_con}

In this paper, we investigated several delay soliton equations, and showed reductions of them lead to delay Painlev\'e equations.
Then, applying the reductions to determinant solutions of delay soliton equations, we derived several determinant solutions of autonomous versions of delay Painlev\'e equations.

In sections \ref{sec_disLV} to \ref{sec_contiTL}, we constructed the Lax pairs and infinite conserved quantities of the delay LV and delay Toda lattice equations and their discrete analogues.
Generalizing these discussions, we can summarize a method for constructing Lax pairs and conserved quantities of delay soliton equations as follows.
\begin{itemize}
\item[1] We first derive a B\"{a}cklund transformation of a discrete analogue of 
a delay soliton equation.
As normal, it can be derived by a reduction of the B\"{a}cklund transformation of the DAGTE~\cite{Hirota1,Miwa1}.
\item[2] By defining $\psi$ as the division of the two solutions of the B\"{a}cklund transformation, the linear problem of $\psi$ and the Lax pair of the discrete-time delay soliton equation is obtained.
\item[3] Because the compatibility condition of the Lax equations includes the delay $\alpha$,
%conserved quantities cannot be obtained by calculating the traces of the Lax operators. Thus
we define a new linear operator $\bar{\mathcal{N}}$ such as (\ref{dlydisLVop0}) in order to remove the delay $\alpha$ from the compatibility condition.
\item[4] Calculating the traces of the powers of $\bar{\mathcal{N}}$, we obtain infinite conserved quantities of the discrete-time delay soliton equation.
\item[5] As the delay-differential limits of the above 1 to 4, we obtain a B\"{a}cklund transformation, Lax pair, and conserved quantities of the continuous delay soliton equation.
We remark that applying the delay-differential limit yields complicated results of conserved quantities since the parameter $\alpha$, which represents the number of operators consisting of $\bar{\mathcal{N}}$, goes to $\infty$.
It is difficult to calculate them explicitly without any conditions, however they can be represented simply by using the time-ordered product.
\end{itemize}

%The above steps 1, 2 and 4 are based on the discussion in~\cite{Hirota2}.

We note that another version of the delay LV equation and its discrete analogue (which are different from the ones addressed in this paper) were proposed before by Sekiguchi \textit{et al}~\cite{Sekiguchi} with their $LR$ formulations.
In future studies, we should reveal a relationship between the Lax pairs of these delay LV equations.

The discussions in this paper can be applied to other delay soliton equations, such as the delay sine-Gordon equation~\cite{Nakata1}, and delay KdV equation~\cite{Nakata2} similarly.
In future studies, we intend to construct their Lax pairs and conserved quantities and reveal relationships with Painlev\'e equations.

In section \ref{sec_reduction}, we showed the $2$-periodic reductions of the delay LV and Toda lattice equations lead to the already known delay Painlev\'e I\hspace{-1pt}I and I\hspace{-1pt}I\hspace{-1pt}I equations.
In addition, the same reductions of the discrete-time delay LV and Toda lattice equations yield equations (\ref{dlydisp2_nl}) and (\ref{dlydisp3_nl}), which lead to the already known discrete and delay Painlev\'e I\hspace{-1pt}I and I\hspace{-1pt}I\hspace{-1pt}I equations as shown in figures \ref{fig_LV} and \ref{fig_TL}.
Thus, equations (\ref{dlydisp2_nl}) and (\ref{dlydisp3_nl}) can be considered as higher order analogues of the discrete Painlev\'e I\hspace{-1pt}I and I\hspace{-1pt}I\hspace{-1pt}I equations respectively.
\begin{figure}
    \includegraphics[width=15cm]
    {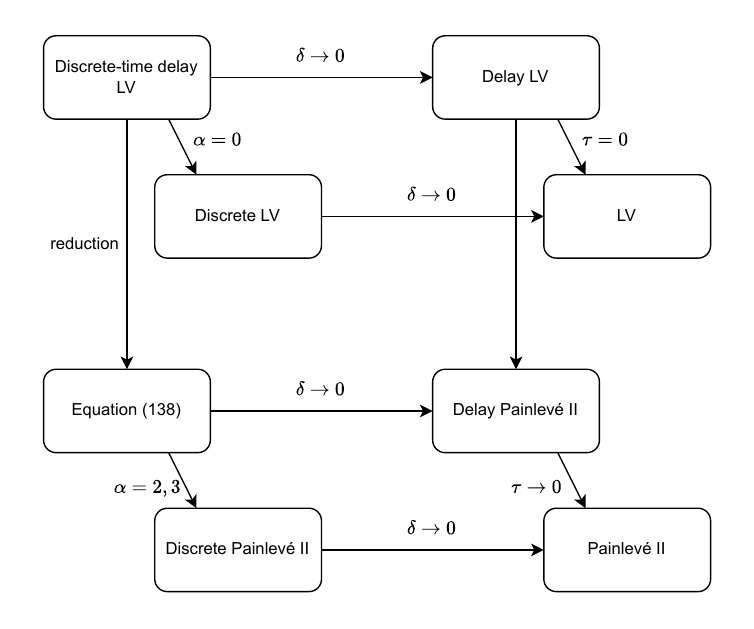}
    \caption{\label{fig_LV}
    Relationship diagram of the LV and Painlev\'e I\hspace{-1pt}I families.}
\end{figure}
\begin{figure}
    \includegraphics[width=15cm]
    {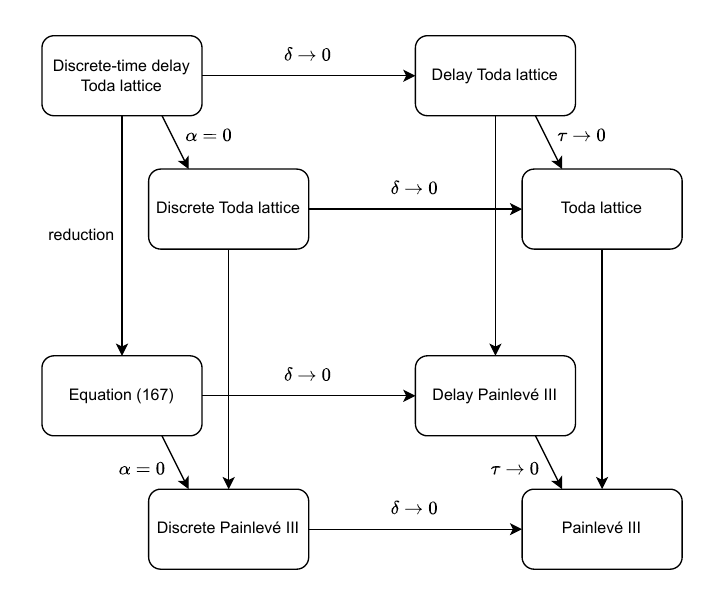}
    \caption{\label{fig_TL}
    Relationship diagram of the Toda lattice and Painlev\'e I\hspace{-1pt}I\hspace{-1pt}I families.}
\end{figure}

The connection between delay soliton equations and delay Painlev\'e equations obtained in section \ref{sec_reduction} allows us to apply the results of delay soliton equations to delay Painlev\'e equations.
In section \ref{sec_detsol}, using the above reductions, we derived the $N$-soliton-type determinant solutions of the autonomous versions of the delay Painlev\'e I\hspace{-1pt}I and I\hspace{-1pt}I\hspace{-1pt}I equations, extending the 3-soliton-type solutions obtained in~\cite{Berntson}.
The reason why the autonomous ordinary differential equations have such arbitrarily large determinant solutions is because they contain delay parameters.

We also constructed the Casorati determinant solution of a type of higher order analogue of the multiplicative discrete Painlev\'e I\hspace{-1pt}I equation (\ref{dlydisp2_alt_nl_1_particular}) and (\ref{dlydisp2_alt_nl_2_particular}) by direct calculation.
In this higher order analogue, the iteration in the column direction in the determinant is generalised from increasing by 2 to increasing by $\alpha$.
We expect higher order analogues of the discrete Painlev\'e I\hspace{-1pt}I and I\hspace{-1pt}I\hspace{-1pt}I equations (\ref{dlydisp2_nl}) and (\ref{dlydisp3_nl}) to have the Casorati determinant solutions similarly.
%In future studies, we intend to address these problems.
In addition, the method of constructing Lax pairs proposed in this paper can be applied to delay Painlev\'e equations.
In our forthcoming paper, we discuss these issues.

\end{section}

\ack
This work was partially supported by JSPS KAKENHI Grant Numbers 22K03441, 22H01136 and Waseda University Grants for Special Research Projects.

\appendix

\begin{section}{Derivations of (\ref{dlyLVop2}), (\ref{dlyLV_E_2}) and (\ref{dlyLV_E})}
\label{sec_derivation}

First, we establish the following relations:
\begin{eqnarray}
    \lim_{\delta\to0}\bar{\mathcal{M}}(m)=\mathcal{M}(t)\,,\label{limitM}\\
    \lim_{\delta\to0}\bar{\mathcal{L}}_{2}(m)=\mathcal{L}_{2}(t)\,.\label{limitL2}
\end{eqnarray}
Using (\ref{dlydisLVMbar}), (\ref{dlydisLVLaxMatrixL1}) in order and setting $\mu=1/\delta$, we derive
\begin{eqnarray}
    \fl\bar{\mathcal{M}}(m)
    &=&\bar{\mathcal{L}}_{1}(m)-\frac{1}{\delta}I\nonumber\\
    \fl&=&\left(
    \begin{array}{ccccc}
    (R_{0}^{m}-1)/\delta & 0             & 0             & \cdots & 1 \\
    1             & (R_{1}^{m}-1)/\delta & 0             & \cdots & 0 \\
    0             & 1             & (R_{2}^{m}-1)/\delta & \cdots & 0 \\
    \vdots        & \vdots        & \ddots        & \ddots & \vdots \\
    0             & 0             & \cdots        & 1      & (R_{N-1}^{m}-1)/\delta \\
    \end{array}
    \right)\,.
\end{eqnarray}
From (\ref{dlydisLVRS}) and (\ref{dlyLVT}) in order, we obtain the following relation in the small limit of $\delta$.
\begin{eqnarray}
    \fl\lim_{\delta\to0}\bar{\mathcal{M}}(m)
    &=&\left(
    \begin{array}{ccccc}
    T_{0}(t) & 0        & 0        & \cdots & 1 \\
    1        & T_{1}(t) & 0        & \cdots & 0 \\
    0        & 1        & T_{2}(t) & \cdots & 0 \\
    \vdots   & \vdots   & \ddots   & \ddots & \vdots \\
    0        & 0        & \cdots   & 1      & T_{N-1}(t) \\
    \end{array}
    \right)\,.
\end{eqnarray}
From (\ref{dlyLVmatrixM}), we obtain (\ref{limitM}).
Following the same approach, using (\ref{dlydisLVLaxMatrixL2}), (\ref{dlydisLVRS}), (\ref{dlyLVtrans}), (\ref{dlyLVmatrixL2}) in order and $\mu=1/\delta$, we obtain (\ref{limitL2}).
    
Next, we apply the delay-differential limit (\ref{dlydiffLimitLV}) to $\bar{\mathcal{N}}^{*}(m)$ in (\ref{dlydisLVop2}).
\begin{eqnarray*}
    \fl \bar{\mathcal{N}}^{*}(m)
    &=&(I+\delta \bar{\mathcal{M}}(m))(I+\delta \bar{\mathcal{M}}(m+1))\cdots(I+\delta \bar{\mathcal{M}}(m+\alpha-1))\bar{\mathcal{L}}_{2}(m)^{-1}\nonumber\\
    \fl&=&\left(I+\sum_{0\leq i_{1}\leq\alpha-1}\delta \bar{\mathcal{M}}(m+i_{1})+\sum_{0\leq i_{1}<i_{2}\leq\alpha-1}\delta^2\bar{\mathcal{M}}(m+i_{1})\bar{\mathcal{M}}(m+i_{2})\right.\nonumber\\
    \fl&&\left.+\sum_{0\leq i_{1}<i_{2}<i_{3}\leq\alpha-1}\delta^3\bar{\mathcal{M}}(m+i_{1})\bar{\mathcal{M}}(m+i_{2})\bar{\mathcal{M}}(m+i_{3})+\cdots\right)\bar{\mathcal{L}}_{2}(m)^{-1}\nonumber\\
    \fl&\longrightarrow&\left(I+\int_{0}^{2\tau}dx_{1}\mathcal{M}(t+x_{1})+\int_{0}^{2\tau}dx_{2}\int_{0}^{x_{2}}dx_{1}\mathcal{M}(t+x_{1})\mathcal{M}(t+x_{2})\right.\nonumber\\
    \fl&&\left.+\int_{0}^{2\tau}dx_{3}\int_{0}^{x_{3}}dx_{2}\int_{0}^{x_{2}}dx_{1}\mathcal{M}(t+x_{1})\mathcal{M}(t+x_{2})\mathcal{M}(t+x_{3})+\cdots\right)\mathcal{L}_{2}(t)^{-1}\,.
\end{eqnarray*}
Denoting the delay-differential limit of $\bar{\mathcal{N}}^{*}(m)$ as $\mathcal{N}^{*}(t)$, the above expression can be rewritten as
\begin{eqnarray}
    \fl\mathcal{N}^{*}(t)&=&E(t)\mathcal{L}_{2}(t)^{-1}\,,\\
    \fl E(t)&=&I+\int_{0}^{2\tau}dx_{1} \mathcal{M}(t+x_{1})+\int_{0}^{2\tau}dx_{2}\int_{0}^{x_{2}}dx_{1} \mathcal{M}(t+x_{1})\mathcal{M}(t+x_{2})\nonumber\\
    \fl&&+\int_{0}^{2\tau}dx_{3}\int_{0}^{x_{3}}dx_{2}\int_{0}^{x_{2}}dx_{1} \mathcal{M}(t+x_{1})\mathcal{M}(t+x_{2})\mathcal{M}(t+x_{3})+\cdots\,.\label{dlyLV_E_2_appendix}
\end{eqnarray}
These are exactly (\ref{dlyLVop2}) and (\ref{dlyLV_E_2}).

Finally, we calculate (\ref{dlyLV_E_2_appendix}) as follows.
\begin{eqnarray}
    \fl E(t)&=&I+\int_{0}^{2\tau}dx_{1} \mathcal{M}(t+x_{1})+\int_{0}^{2\tau}dx_{2}\int_{0}^{x_{2}}dx_{1} \mathcal{M}(t+x_{1})\mathcal{M}(t+x_{2})\nonumber\\
    \fl&&+\int_{0}^{2\tau}dx_{3}\int_{0}^{x_{3}}dx_{2}\int_{0}^{x_{2}}dx_{1} \mathcal{M}(t+x_{1})\mathcal{M}(t+x_{2})\mathcal{M}(t+x_{3})+\cdots\nonumber\\
    \fl&=&I+\int_{0\leq x_{1}\leq2\tau}dx_{1} \mathcal{M}(t+x_{1})+\int\!\!\!\int_{0\leq x_{1}\leq x_{2}\leq2\tau}dx_{1}dx_{2} \mathcal{M}(t+x_{1})\mathcal{M}(t+x_{2})\nonumber\\
    \fl&&+\int\!\!\!\int\!\!\!\int_{0\leq x_{1}\leq x_{2}\leq x_{3}\leq2\tau}dx_{1}dx_{2}dx_{3} \mathcal{M}(t+x_{1})\mathcal{M}(t+x_{2})\mathcal{M}(t+x_{3})+\cdots\nonumber\\
    \fl&=&T\left(I+\int_{0\leq x_{1}\leq2\tau}dx_{1} \mathcal{M}(t+x_{1})+\frac{1}{2!}\int\!\!\!\int_{0\leq x_{1}\leq2\tau,\ 0\leq x_{2}\leq2\tau}dx_{1}dx_{2} \mathcal{M}(t+x_{1})\mathcal{M}(t+x_{2})\right.\nonumber\\
    \fl&&\left.+\frac{1}{3!}\int\!\!\!\int\!\!\!\int_{0\leq x_{1},x_{2},x_{3}\leq2\tau}dx_{1}dx_{2}dx_{3} \mathcal{M}(t+x_{1})\mathcal{M}(t+x_{2})\mathcal{M}(t+x_{3})+\cdots\right)\nonumber\\
    \fl&=&T\left(I+\int_{0}^{2\tau}ds \mathcal{M}(t+s)+\frac{1}{2!}\left(\int_{0}^{2\tau}ds \mathcal{M}(t+s)\right)^2+\frac{1}{3!}\left(\int_{0}^{2\tau}ds \mathcal{M}(t+s)\right)^3+\cdots\right)\nonumber\\
    \fl&=&T\left(\exp\left(\int_{0}^{2\tau}\mathcal{M}(t+s)ds\right)\right)\,.\label{dlyLV_E_3_appendix}
\end{eqnarray}
Thus, we have derived (\ref{dlyLV_E}).

\begin{rem}
As mentioned in (\ref{time-ordered_product}), $T$ is the time-ordered product defined by 
\begin{eqnarray}
    T\left(A(t_{1})A(t_{2})\cdots A(t_{i})\right)=A(t_{\sigma(1)})A(t_{\sigma(2)})\cdots A(t_{\sigma(i)})\,,
\end{eqnarray}
where $\sigma$ is a permutation that satisfies $t_{\sigma(1)}\leq t_{\sigma(2)}\leq\cdots\leq t_{\sigma(i)}$.
We assume that $T$ commutes with integration.
Under this assumption, the following property holds and is used in the above calculation (\ref{dlyLV_E_3_appendix}).
\begin{eqnarray*}
    \fl&&\int\!\!\!\int_{0\leq x_{1}\leq x_{2}\leq2\tau}dx_{1}dx_{2} \mathcal{M}(t+x_{1})\mathcal{M}(t+x_{2})\\
    \fl&=&\int\!\!\!\int_{0\leq x_{1}\leq x_{2}\leq2\tau}dx_{1}dx_{2} \left(\frac{1}{2}\mathcal{M}(t+x_{1})\mathcal{M}(t+x_{2})+\frac{1}{2}\mathcal{M}(t+x_{1})\mathcal{M}(t+x_{2})\right)\\
    \fl&=&\int\!\!\!\int_{0\leq x_{1}\leq x_{2}\leq2\tau}dx_{1}dx_{2} \left(\frac{1}{2}T(\mathcal{M}(t+x_{1})\mathcal{M}(t+x_{2}))+\frac{1}{2}T(\mathcal{M}(t+x_{2})\mathcal{M}(t+x_{1}))\right)\\
    \fl&=&T\left(\frac{1}{2}\int\!\!\!\int_{0\leq x_{1}\leq x_{2}\leq2\tau}dx_{1}dx_{2} \left(\mathcal{M}(t+x_{1})\mathcal{M}(t+x_{2})+\mathcal{M}(t+x_{2})\mathcal{M}(t+x_{1})\right)\right)\\
    \fl&=&T\left(\frac{1}{2!}\int\!\!\!\int_{0\leq x_{1}\leq2\tau,\ 0\leq x_{2}\leq2\tau}dx_{1}dx_{2} \mathcal{M}(t+x_{1})\mathcal{M}(t+x_{2})\right)\,.
\end{eqnarray*}
We note that $\mathcal{M}(t + x_1)$ and $\mathcal{M}(t + x_2)$ are reordered in the second equality, and that $T$ is interchanged with the integral in the third equality.
\end{rem}

\end{section}

\begin{section}{Proof of Theorem \ref{thm_detsol}}
\label{sec_proof}

We show the proof of theorem \ref{thm_detsol}.
What we need to do is reduce (\ref{dlydisp2_alt_1_particular}) and (\ref{dlydisp2_alt_2_particular}) to Pl\"ucker relations by a similar approach to that in~\cite{Kajiwara}.

We first introduce a notation
\begin{equation}
    \tau_{N}(n)=\left|(0),(-1),(-2),\cdots,(-N+2),(-N+1)\right|\,,
\end{equation}
where
\begin{equation}
    (j)=\left(
    \begin{array}{c}
        f_{n+j}\\
        f_{n+\alpha+j}\\
        f_{n+2\alpha+j}\\
        \vdots
    \end{array}
    \right)\,.
\end{equation}
The size of $(j)$ is defined as appropriate case by case.
By adding the $i$-th row of (\ref{dlydisp2_alt_tau}) multiplied by $-\delta$ to the $(i+1)$-th row from $i=N-1$ to $i=1$, and using (\ref{dlydisp2_alt_disper}), we obtain
\begin{eqnarray*}
    \fl\tau_{N}(n)=\delta^{N-1}(-1)^{\alpha(N-1)}\\
    \fl\times\left|
    \begin{array}{cccc}
        f_{n}&
        f_{n+\alpha}&\cdots&
        f_{n+(N-1)\alpha}\\
        q^{2(\alpha-1)(n-\alpha+1)}f_{n-\alpha}&
        q^{2(\alpha-1)(n+1)}f_{n}&\cdots&
        q^{2(\alpha-1)(n+(N-2)\alpha+1)}f_{n+(N-2)\alpha}\\
        \vdots&\vdots&\ddots&\vdots\\
        q^{2(\alpha-1)(n-\alpha-N+3)}f_{n-\alpha-N+2}&
        q^{2(\alpha-1)(n-N+3)}f_{n-N+2}&\cdots&
        q^{2(\alpha-1)(n+(N-2)(\alpha-1)+1)}f_{n+(N-2)(\alpha-1)}\\
    \end{array}
    \right|\\ \\
    \fl=\delta^{N-1}(-1)^{\alpha(N-1)}q^{2(\alpha-1)n(N-1)-2(\alpha-1)(\alpha-2)(N-1)+(\alpha-1)^2N(N-1)}\\
    \fl\times\left|
    \begin{array}{cccc}
        f_{n}&
        q^{-2(\alpha-1)\alpha}f_{n+\alpha}&\cdots&
        q^{-2(\alpha-1)\alpha(N-1)}f_{n+(N-1)\alpha}\\
        f_{n-\alpha}&
        f_{n}&\cdots&
        f_{n+(N-2)\alpha}\\
        \vdots&\vdots&\ddots&\vdots\\
        f_{n-\alpha-N+2}&
        f_{n-N+2}&\cdots&
        f_{n+(N-2)(\alpha-1)}\\
    \end{array}
    \right|\\ \\
    \fl=\delta^{N-1}(-1)^{\alpha(N-1)}q^{2(\alpha-1)n(N-1)-2(\alpha-1)(\alpha-2)(N-1)+(\alpha-1)^2N(N-1)}\\
    \fl\times\left|(0)',(-\alpha),(-\alpha-1),\cdots,(-\alpha-N+3),(-\alpha-N+2)\right|\,,
\end{eqnarray*}
where
\begin{equation}
    (j)'=\left(
    \begin{array}{c}
        f_{n+j}\\
        q^{-2(\alpha-1)\alpha}f_{n+\alpha+j}\\
        q^{-2(\alpha-1)\alpha\times2}f_{n+2\alpha+j}\\
        \vdots
    \end{array}
    \right)\,.
\end{equation}
The size of $(j)'$ is defined as appropriate case by case.
Next, continuing the above calculation, we add the $2$nd row multiplied by $(-1)^{\alpha}q^{2(\alpha-1)(n-\alpha+1)}$ to the $1$st row, and use (\ref{dlydisp2_alt_disper}).
Then we obtain
\begin{eqnarray*}
    \fl\tau_{N}(n)=\delta^{N-2}(-1)^{\alpha(N-1)}q^{2(\alpha-1)n(N-1)-2(\alpha-1)(\alpha-2)(N-1)+(\alpha-1)^2N(N-1)}\\
    \fl\times\left|
    \begin{array}{cccc}
        f_{n-1}&
        q^{-2(\alpha-1)\alpha}f_{n+\alpha-1}&\cdots&
        q^{-2(\alpha-1)\alpha(N-1)}f_{n+(N-1)\alpha-1}\\
        f_{n-\alpha}&
        f_{n}&\cdots&
        f_{n+(N-2)\alpha}\\
        \vdots&\vdots&\ddots&\vdots\\
        f_{n-\alpha-N+2}&
        f_{n-N+2}&\cdots&
        f_{n+(N-2)(\alpha-1)}\\
    \end{array}
    \right|\\ \\
    \fl=\delta^{N-2}(-1)^{\alpha(N-1)}q^{2(\alpha-1)n(N-1)-2(\alpha-1)(\alpha-2)(N-1)+(\alpha-1)^2N(N-1)}\\
    \fl\times\left|(-1)',(-\alpha),(-\alpha-1),\cdots,(-\alpha-N+3),(-\alpha-N+2)\right|\,.
\end{eqnarray*}

According to the above calculations, we obtain the following formulas:
\begin{eqnarray}
    \label{formula1}
    \fl\left\{
    \begin{array}{lll}
    F_{n}&=&\tau_{N}(n)\\\\
    &=&\left|(0),(-1),(-2),\cdots,(-N+2),(-N+1),\phi\right|\\\\
    &=&\delta^{N-1}(-1)^{\alpha(N-1)}q^{2(\alpha-1)n(N-1)-2(\alpha-1)(\alpha-2)(N-1)+(\alpha-1)^2N(N-1)}\\
    &&\times\left|(0)',(-\alpha),(-\alpha-1),\cdots,(-\alpha-N+3),(-\alpha-N+2),\phi\right|\\\\
    &=&\delta^{N-2}(-1)^{\alpha(N-1)}q^{2(\alpha-1)n(N-1)-2(\alpha-1)(\alpha-2)(N-1)+(\alpha-1)^2N(N-1)}\\
    &&\times\left|(-1)',(-\alpha),(-\alpha-1),\cdots,(-\alpha-N+3),(-\alpha-N+2),\phi\right|\,,
    \end{array}
    \right.\\\nonumber\\
    \label{formula2}
    \fl\left\{
    \begin{array}{lll}
    F_{n}&=&\tau_{N}(n)\\\\
    &=&\left|\psi,(-\alpha),(-\alpha-1),(-\alpha-2),\cdots,(-\alpha-N+2),(-\alpha-N+1)\right|\\\\
    &=&\delta^{N-1}(-1)^{\alpha(N-1)}q^{2(\alpha-1)n(N-1)-2(\alpha-1)(\alpha-2)(N-1)+(\alpha-1)^2N(N-1)+2(\alpha-1)\alpha}\\
    &&\times\left|\psi,(-\alpha)',(-2\alpha),(-2\alpha-1),\cdots,(-2\alpha-N+3),(-2\alpha-N+2)\right|\\\\
    &=&\delta^{N-2}(-1)^{\alpha(N-1)}q^{2(\alpha-1)n(N-1)-2(\alpha-1)(\alpha-2)(N-1)+(\alpha-1)^2N(N-1)+2(\alpha-1)\alpha}\\
    &&\times\left|\psi,(-\alpha-1)',(-2\alpha),(-2\alpha-1),\cdots,(-2\alpha-N+3),(-2\alpha-N+2)\right|\,,
    \end{array}
    \right.\\\nonumber\\
    \label{formula3}
    \fl\left\{
    \begin{array}{lll}
    G_{n}&=&\tau_{N+1}(n)\\\\
    &=&\left|(0),(-1),(-2),\cdots,(-N+1),(-N)\right|\\\\
    &=&\delta^{N}(-1)^{\alpha N}q^{2(\alpha-1)nN-2(\alpha-1)(\alpha-2)N+(\alpha-1)^2(N+1)N}\\
    &&\times\left|(0)',(-\alpha),(-\alpha-1),\cdots,(-\alpha-N+2),(-\alpha-N+1)\right|\\\\
    &=&\delta^{N-1}(-1)^{\alpha N}q^{2(\alpha-1)nN-2(\alpha-1)(\alpha-2)N+(\alpha-1)^2(N+1)N}\\
    &&\times\left|(-1)',(-\alpha),(-\alpha-1),\cdots,(-\alpha-N+2),(-\alpha-N+1)\right|\,,
    \end{array}
    \right.
\end{eqnarray}
where the $(N+1)$-dimensional column vectors $\phi,\psi$ are defined by
\begin{equation}
    \phi=\left(
    \begin{array}{c}
        0\\
        \vdots\\
        0\\
        1
    \end{array}
    \right)\,,\qquad
    \psi=\left(
    \begin{array}{c}
        1\\
        0\\
        \vdots\\
        0
    \end{array}
    \right)\,.
\end{equation}

Now, using (\ref{formula1}) and (\ref{formula3}), the bilinear equation (\ref{dlydisp2_alt_1_particular}) reduces to the following Pl\"ucker relation:
\begin{eqnarray}
    &&\left|(-1),(-2),\cdots,(-N),\phi\right|
    \times\left|(\alpha-1)',(0),\cdots,(-N+1)\right|\nonumber\\
    &-&\left|(0),(-1),\cdots,(-N+1),\phi\right|
    \times\left|(\alpha-1)',(-1),\cdots,(-N)\right|\nonumber\\
    &+&\left|(\alpha-1)',(-1),\cdots,(-N+1),\phi\right|
    \times\left|(0),(-1),\cdots,(-N)\right|\nonumber\\
    &=&0\,.\label{plucker1}
\end{eqnarray}
Using (\ref{formula2}) and (\ref{formula3}), the bilinear equation (\ref{dlydisp2_alt_2_particular}) reduces to the following Pl\"ucker relation:
\begin{eqnarray}
    \fl
    &&\left|\psi,(-\alpha),(-\alpha-1),\cdots,(-\alpha-N+1)\right|
    \times\left|(0)',(-\alpha+1),\cdots,(-\alpha-N+2)\right|\nonumber\\
    \fl
    &-&\left|\psi,(-\alpha+1),(-\alpha),\cdots,(-\alpha-N+2)\right|
    \times\left|(0)',(-\alpha),\cdots,(-\alpha-N+1)\right|\nonumber\\
    \fl
    &+&\left|\psi,(0)',(-\alpha),\cdots,(-\alpha-N+2)\right|
    \times\left|(-\alpha+1),(-\alpha),\cdots,(-\alpha-N+1)\right|\nonumber\\
    \fl
    &=&0\,.\label{plucker2}
\end{eqnarray}
Theorem \ref{thm_detsol} has been now proved.

Note that the Pl\"ucker relation (\ref{plucker1}) is obtained by applying the Laplace expansion to the left-hand side of the identity
\begin{eqnarray}
    \fl\left|
    \begin{array}{cc|c|c|cc}
    (\alpha-1)' & (0) & (-1)\cdots(-N+1) & \varnothing & (-N) & \phi \\
    \hline
    (\alpha-1)' & (0) & \varnothing & (-1)\cdots(-N+1) & (-N) & \phi
    \end{array}
    \right|
    =0\,.
\end{eqnarray}
Similarly, the Pl\"ucker relation (\ref{plucker2}) is obtained by the identity
\begin{eqnarray}
    \fl\left|
    \begin{array}{ccc|c|c|c}
    \psi & (0)' & (-\alpha+1) & (-\alpha)\cdots(-\alpha-N+2) & \varnothing & (-\alpha-N+1) \\
    \hline
    \psi & (0)' & (-\alpha+1) & \varnothing & (-\alpha)\cdots(-\alpha-N+2) & (-\alpha-N+1)
    \end{array}
    \right|\nonumber\\
    =0\,.
\end{eqnarray}

\end{section}

\end{document}